\documentclass[aps,prl,superscriptaddress,longbibliography,twocolumn]{revtex4-1}

\usepackage{lmodern}

\usepackage[latin9]{inputenc}
\usepackage{amsmath}
\usepackage{amssymb}
\usepackage{graphicx}
\usepackage{epstopdf}
\usepackage{color}
\usepackage{enumerate}
\usepackage{ulem}

\begin{document}
\title{Helical quantum Hall phase in graphene on SrTiO$_3$}

\author{Louis Veyrat}
\author{Corentin D\'{e}prez}
\author{Alexis Coissard}
\affiliation{Univ. Grenoble Alpes, CNRS, Grenoble INP, Institut N\'{e}el, 38000 Grenoble, France}
\author{Xiaoxi Li}
\affiliation{Shenyang National Laboratory for Materials Science, Institute of Metal Research, Chinese Academy of Sciences, Shenyang 110016, P. R. China}
\affiliation{School of Material Science and Engineering, University of Science and Technology of China, Anhui 230026, P. R. China}
\affiliation{State Key Laboratory of Quantum Optics and Quantum Optics Devices, Institute of Opto-Electronics, Shanxi University, Taiyuan 030006, P. R. China}
\author{Fr\'{e}d\'{e}ric Gay}
\affiliation{Univ. Grenoble Alpes, CNRS, Grenoble INP, Institut N\'{e}el, 38000 Grenoble, France}
\author{Kenji Watanabe}
\author{Takashi Taniguchi}
\affiliation{National Institute for Materials Science, 1-1 Namiki, Tsukuba 306-0044, Japan}
\author{Zheng Han}
\affiliation{Shenyang National Laboratory for Materials Science, Institute of Metal Research, Chinese Academy of Sciences, Shenyang 110016, P. R. China}
\affiliation{School of Material Science and Engineering, University of Science and Technology of China, Anhui 230026, P. R. China}
\affiliation{State Key Laboratory of Quantum Optics and Quantum Optics Devices, Institute of Opto-Electronics, Shanxi University, Taiyuan 030006, P. R. China}
\author{Benjamin A. Piot}
\affiliation{Laboratoire National des Champs Magn\'{e}tiques Intenses, LNCMI-CNRS-UGA-UPS-INSA-EMFL,
F-38042 Grenoble, France}
\author{Hermann Sellier}
\author{Benjamin Sac\'{e}p\'{e}}
\email{benjamin.sacepe@neel.cnrs.fr}
\affiliation{Univ. Grenoble Alpes, CNRS, Grenoble INP, Institut N\'{e}el, 38000 Grenoble, France}

\date{\today}

\begin{abstract}
The ground state of charge neutral graphene under perpendicular magnetic field was predicted to be a quantum Hall topological insulator with a ferromagnetic order and spin-filtered, helical edge channels. In most experiments, however, an otherwise insulating state is observed and is accounted for by lattice-scale interactions that promote a broken-symmetry state with gapped bulk and edge excitations. We tuned the ground state of the graphene zeroth Landau level to the topological phase via a suitable screening of the Coulomb interaction with a SrTiO$_3$ high-$k$ dielectric substrate. We observed robust helical edge transport emerging at a magnetic field as low as 1 tesla and withstanding temperatures up to 110 kelvins over micron-long distances. This new and versatile graphene platform opens new avenues for spintronics and topological quantum computation.
\end{abstract}

\maketitle

There is a variety of topological phases that are classified by their dimensionality, symmetries and topological invariants~\cite{Hasan10,Qi2011}. They all share the remarkable property that the topological bulk gap closes at every interfaces with vacuum or a trivial insulator, forming conductive edge states with peculiar transport and spin properties. The quantum Hall effect that arises in two-dimensional (2D) electron systems subjected to a perpendicular magnetic field, $B$, stands out as a paradigmatic example characterized by the Chern number that quantizes the Hall conductivity and counts the number of chiral, one-dimensional edge channels. The singular aspect of quantum Hall systems compared to time-reversal symmetric topological insulators  (TIs) lies in the pivotal role of Coulomb interaction between electrons that can induce a wealth of strongly correlated, symmetry or topologically-protected phases, ubiquitously observed in various experimental systems~\cite{Tsui82,Willett87,Du09,Bolotin09,Maher13,Young14,Ki14,Falson15,Zibrov18,Kim19}. 
 
 \begin{figure*}[t!]
 	\includegraphics[width=0.8\linewidth]{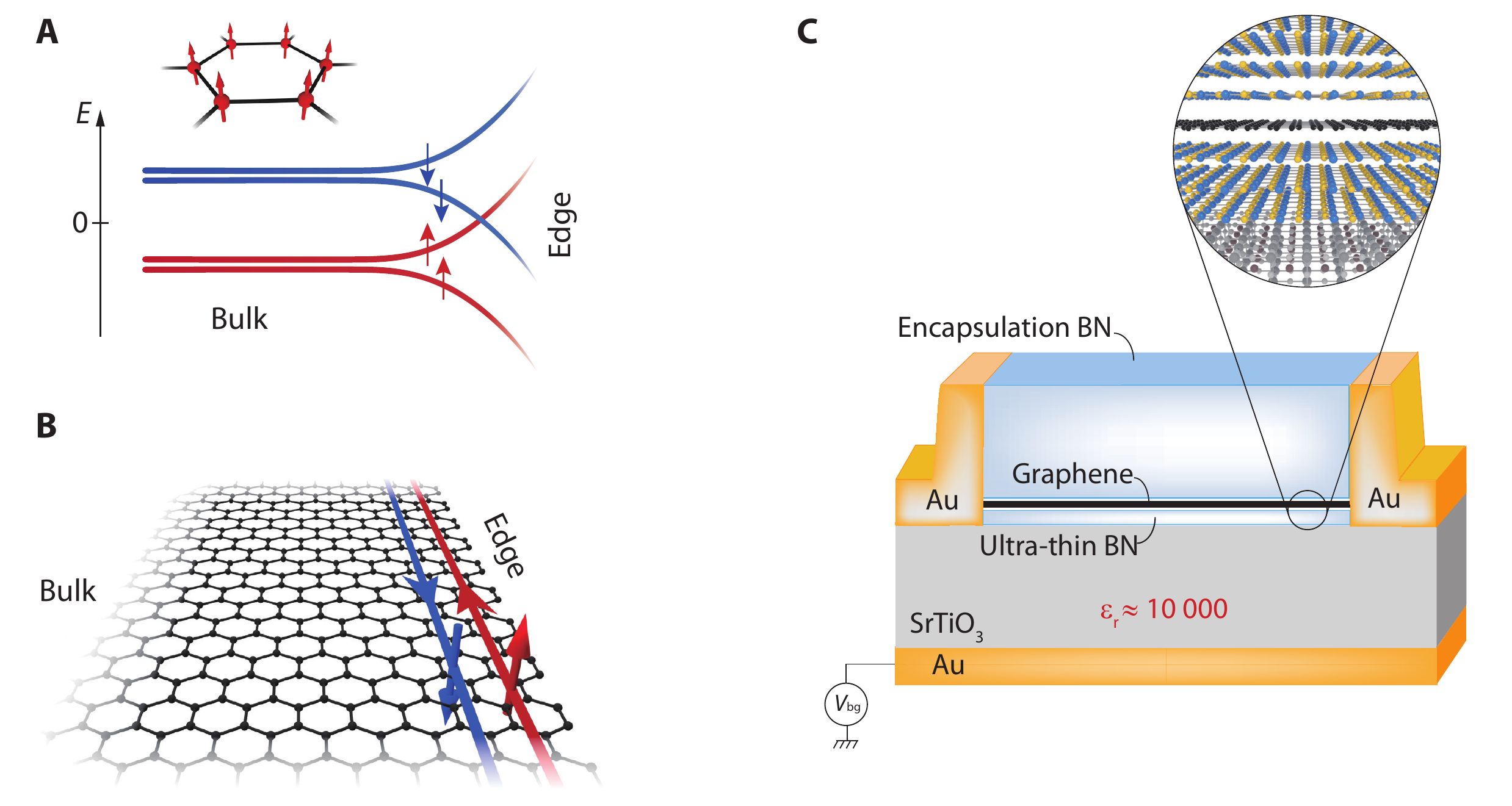} 
 	\caption{\textbf{Spin-polarized ferromagnetic phase in graphene on high-$k$ dielectric.} (\textbf{A}) In the ferromagnetic phase of charge neutral graphene, the broken-symmetry state of the half-filled zeroth Landau level is spin-polarized and occupies both sublattices of the honeycomb lattice as shown in the top inset.  The edge dispersion results from linear combinations of the bulk isospin states, which disperse as electron-like and hole-like branches, yielding a pair of counter-propagative, spin-filtered helical edge channels at charge neutrality.~\cite{Abanin06,Brey06} (\textbf{B}) Schematics of a graphene lattice with helical edge channels propagating on the crystallographic armchair edge. (\textbf{C}) Schematics of the hBN-encapsulated graphene device placed on a SrTiO$_3$ substrate that serves both as high-$k$ dielectric and back-gate dielectric. Due to the considerable dielectric constant of the SrTiO$_3$ substrate at low temperature and the ultra-thin hBN spacer ($2-5\,$nm thick), Coulomb interaction in the graphene plane is significantly screened, resulting in a modification of the quantum Hall ground state at charge neutrality and the emergence of the ferromagnetic phase with helical edge transport. The inset shows the atomic layers of the hBN-encapsulated graphene van der Waals assembly, and the surface atomic structure of SrTiO$_3$.}
 	\label{Fig1}
 \end{figure*}

 In graphene the immediate consequence of the Coulomb interaction is an instability towards quantum Hall ferromagnetism. 
 Due to exchange interaction, a spontaneous breaking of the SU(4) symmetry splinters the Landau levels into quartets of broken-symmetry states that are polarized in one, or a combination of the spin and valley (pseudospin) degrees of freedom~\cite{Yang06,Nomura06,Young12}.
 
 Central to this phenomenon is the fate of the zeroth Landau level and its quantum Hall ground states. It was early predicted that if the Zeeman spin-splitting (enhanced by exchange interaction) overcomes the valley splitting, a topological inversion between the lowest electron-type and highest hole-type sub-levels should occur~\cite{Abanin06,Fertig06}. 
 At charge neutrality, the ensuing ground state is a quantum Hall ferromagnet with two filled states of identical spin polarization, and an edge dispersion that exhibits two counter-propagating, spin-filtered helical edge channels (Fig.~\ref{Fig1}A and B), similar to those of the quantum spin Hall (QSH) effect in 2D TIs~\cite{konig07,Knez11,Fei17,wu18,Hatsuda18}. 
 Such a spin-polarized ferromagnetic (F) phase belongs to the recently identified new class of interaction-induced TIs with zero Chern number, termed quantum Hall topological insulators~\cite{Kharitonov16} (QHTIs), which arise from a many-body interacting Landau level and can be pictured as two independent copies of quantum Hall systems with opposite chiralities. 
Remarkably, unlike 2D TIs, immunity from back-scattering for the helical edge channels does not rely on the discrete time-reversal symmetry, conspicuously broken here by the magnetic field, but on the continuous U(1) axial rotation symmetry of the spin polarization~\cite{Young14,Kharitonov16}.
 
  \begin{figure*}[t!]
  	\includegraphics[width=0.9\linewidth]{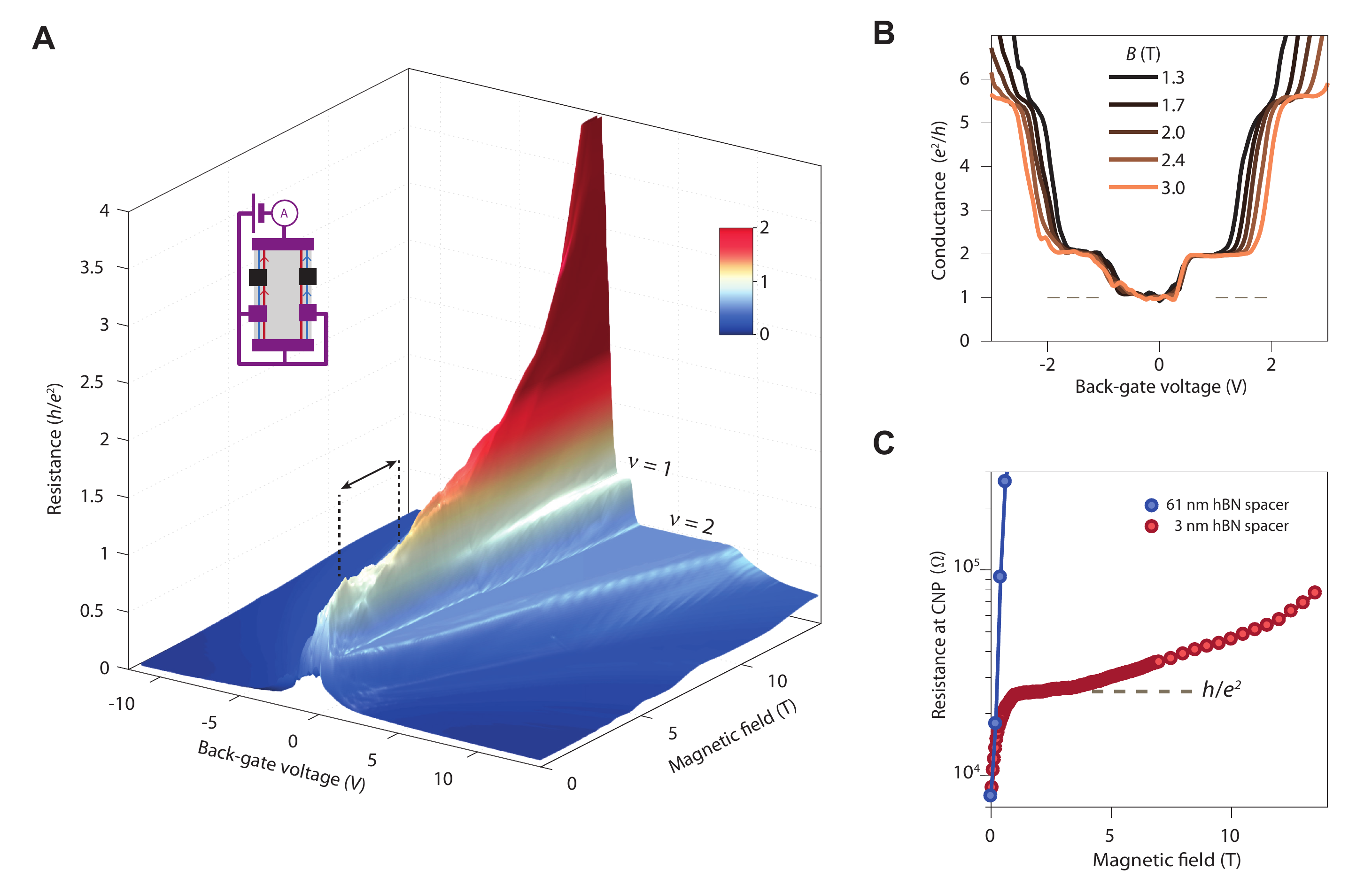} 
  	\caption{\textbf{Low magnetic field quantum spin Hall effect.} (\textbf{A}) Two-terminal resistance $R_{\text{2t}}$ in units of $h/e^2$ of sample BNGrSTO-07 versus magnetic field $B$ and back-gate voltage $V_{\text{bg}}$ measured at $4$~K.  In addition to standard quantum Hall plateaus at filling fraction $\nu=1$ and 2, the resistance exhibits an anomalous plateau around the charge neutrality point between $B=1.5$ and $4$~T, delimited by the black dotted lines and the arrow, which signals the regime of the QSH effect in this sample. The value of the resistance at this plateau is $h/e^2$ and is color-coded in white. The inset schematics indicates the contact configuration. Black contacts are floating. The red and blue arrows on the helical edge channels indicates the direction of the current between contacts. (\textbf{B}) Two-terminal conductance $G_{\text{2t}}=1/R_{\text{2t}}$ in units of $e^2/h$ versus $V_{\text{bg}}$ extracted from (A) at different magnetic fields. The first conductance plateaus of the quantum Hall effect at $2\,e^2/h$ and $6\,e^2/h$ are well defined. The QSH plateau of conductance $e^2/h$ clearly emerges at charge neutrality around $V_{\text{bg}}=0$~V. (\textbf{C}) Resistance at the charge neutrality point (CNP) versus $B$ for sample BNGrSTO-07 (red dots) extracted from (A), and sample BNGrSTO-09 (blue dots).  Whereas the sample with a thick hBN spacer exhibits a strong positive magnetoresistance at low $B$ diverging towards insulation, the sample with the thin hBN spacer shows the QSH plateau that persists up to $\sim 4$~T, followed by a resistance increase at higher $B$.} 
  	\label{Fig2}
  \end{figure*}
	
 The experimental situation is however at odds with this exciting scenario. A strong insulating state is systematically observed on increasing perpendicular magnetic field in charge-neutral, high-mobility graphene devices. The formation of the F phase is presumably hindered by some lattice-scale electron-electron and electron-phonon interaction terms, whose amplitudes and signs can be strongly renormalized by the long-range part of the Coulomb interaction~\cite{Kharitonov12}, favoring various insulating spin- or charge-density-wave orders~\cite{Alicea06,Jung09,Kharitonov12}. Only with a very strong in-plane magnetic field component such that Zeeman energy overcomes the other anisotropic interaction terms does the F phase emerge experimentally~\cite{Young14,Maher13}. Another strategy to engineer a F phase uses small-angle twisted graphene bilayers, with each layer sets in a different quantum Hall state of opposite charge polarity via a displacement field~\cite{Sanchez17}.  Yet, those approaches realized hitherto suffer from either unpractical strong and tilted magnetic field or the complexity of the twisted layers assembly.
 
 In this work we take a novel route to induce the F phase in monolayer graphene in a straightforward fashion. Instead of boosting the Zeeman effect with a strong in-plane field, we mitigate the effects of the lattice-scale interaction terms by a suitable substrate screening of the Coulomb interaction to restore the dominant role of the spin-polarizing terms and induce the F phase. We use a high-$k$ dielectric substrate, the quantum paraelectric SrTiO$_3$ known to exhibit a very large static dielectric constant of the order of $\epsilon \approx 10^4$ at low temperatures~\cite{Sakudo71} (see Fig. S3), which acts both as an electrostatic screening environment and back-gate dielectric~\cite{Couto11}. 
 For an efficient screening of the Coulomb potential, the graphene layer must be sufficiently close to the substrate, with a separation less than the magnetic length $l_B=\sqrt{\hbar/eB}$ ($\hbar$ the reduced Planck constant, $e$ the electron charge), which is the relevant length scale in the quantum Hall regime. High mobility hexagonal boron-nitride (hBN) encapsulated graphene heterostructures~\cite{Wang13} were thus purposely made with an ultra-thin bottom hBN layer with a thickness, $d_{\text{BN}}$,  ranging between $2-5$~nm (see Fig.~\ref{Fig1}C, and~\cite{SM}) inferior to the magnetic length for moderate magnetic field (e.g. $l_B > 8$~nm for $B < 10$~T).

 The emergence of the F phase in such a screened configuration is readily seen in Fig.~\ref{Fig2}A, which displays the two-terminal resistance of a hBN encapsulated graphene device in a six-terminal Hall bar geometry, as a function of the back-gate voltage $V_{\text{bg}}$ and magnetic field. Around the charge neutrality ($V_{\text{bg}}\sim 0$~V), an anomalous resistance plateau develops over a $B$-range from $1.5$ to $4$~T indicated by the two dotted black lines. This plateau reaches the quantum of resistance $h/e^2$, color-coded in white. At $B>5$~T, the resistance departs from $h/e^2$ towards insulation as seen by the red color-coded magneto-resistance peak (see also Fig.~\ref{Fig2}C). 
 
 The unusual nature of this resistance plateau can be captured with the line-cuts of the two-terminal conductance $G_{\text{2t}} = 1/R_{\text{2t}}$ versus $V_{\text{bg}}$ at fixed $B$, see Fig.~\ref{Fig2}B. 
 While standard graphene quantum Hall plateaus at $G_{\text{2t}}= 4\frac{e^2}{h}(N+\frac{1}{2}) = 2\,e^2/h$ and $6\,e^2/h$ for the Landau level indices $N=0$ and $N=1$ are well developed as a function of back-gate voltage, the new plateau at $G_{\text{2t}} =  e^2/h$ is centered at the charge neutrality and does not show any dip at  $V_{\text{bg}} = 0$~V. 
 This behavior is at odds with the usual sequence of broken-symmetry states setting with magnetic field where first the insulating broken-symmetry state opens  at $\nu=0$ with $G_{\text{2t}}=0$, followed at higher field by the plateaus of the broken-symmetry states at filling fraction $\nu=\pm 1$ ~\cite{Nomura06,Young12}.  In Figure~\ref{Fig2}A, the latter states at  $\nu=\pm 1$ arise for $B>6$~T together with the insulating magnetoresistance peak at $\nu=0$, that is, above the field range of the anomalous plateau. Hence, this striking observation of a $h/e^2$ plateau at low magnetic field conspicuously points to a new broken-symmetry state at $\nu=0$. We show in the following that this $h/e^2$ plateau is a direct signature of the QSH effect resulting from the helical edge channels of the F phase.
   \begin{figure*}[t!]
  	\includegraphics[width=0.8\linewidth]{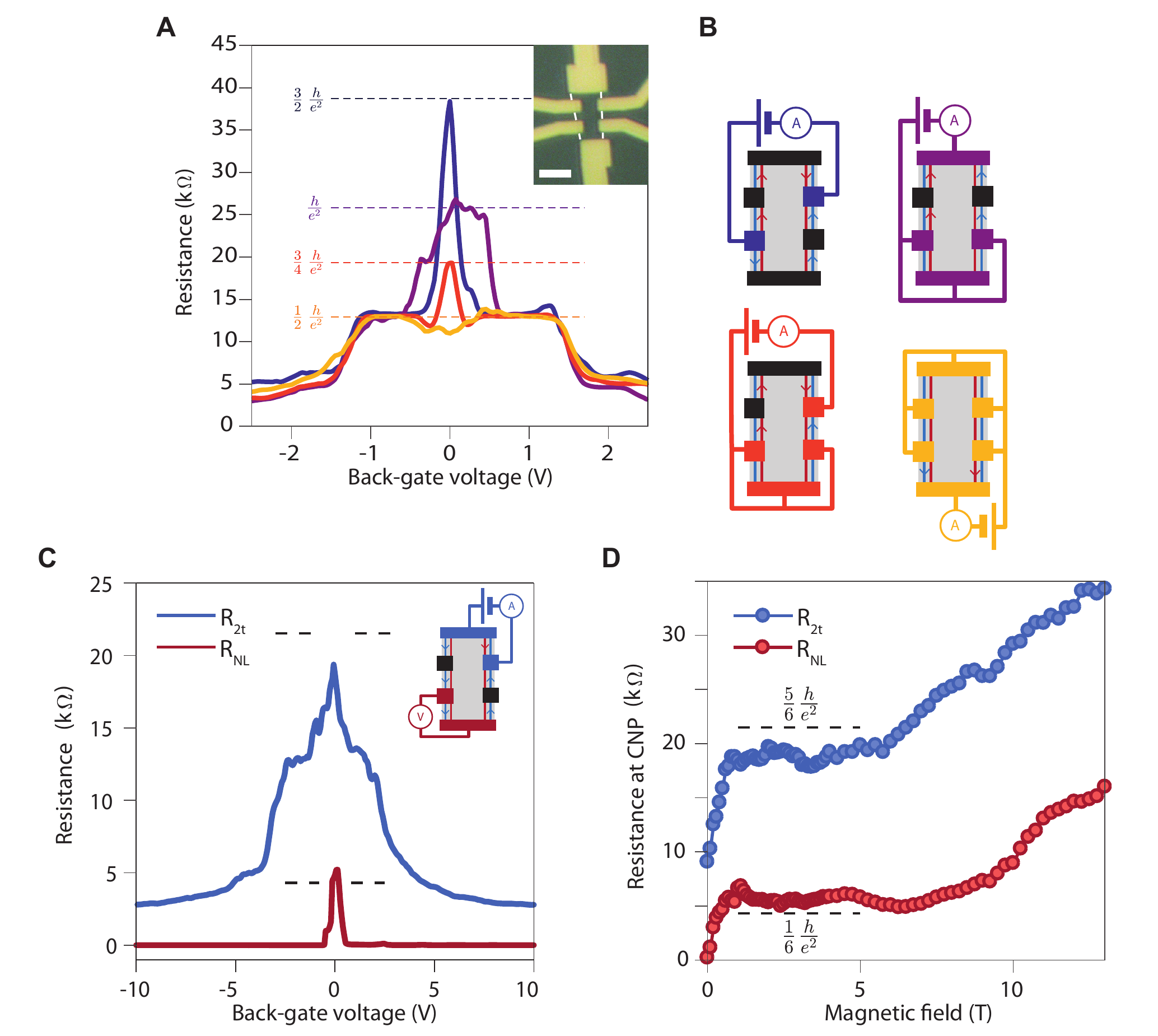} 
  	\caption{\textbf{Non-local helical edge transport.} (\textbf{A}) Two-terminal resistance versus back-gate voltage measured at $2.5$~T for different contact configurations schematized in (B). The inset shows an optical picture of the measured sample BNGrSTO-07. The scale bar is $4\,\mu$m. Each contact configuration yields a resistance at charge neutrality reaching the expected values for helical edge transport, which are indicated with the horizontal dashed lines. (\textbf{B}) Schematics of the measurement configurations. Black contacts are floating. The red and blue arrows on the helical edge channels indicates the direction of the current between contacts. (\textbf{C}) Two-terminal resistance, $R_{\text{2t}}$, in blue and non-local, four-terminal resistance, $R_{\text{NL}}$, in red versus back-gate voltage, in the contact configuration shown in the inset schematics. (\textbf{D}) Resistance at the charge neutrality point (CNP), $V_{\text{bg}}=0$, in the same contact configuration as in (C) versus magnetic field. The helical plateau is observed for both two- and four-terminal resistances between 1~T and about 6~T.}
  	\label{Fig3}
  \end{figure*}
	
 Helical edge transport has unambiguous signatures in multi-terminal device configuration as each ohmic contact acts as a source of back-scattering for the counter-propagating helical edge channels with opposite spin-polarization~\cite{Roth09}. An edge section between two contacts is indeed an ideal helical quantum conductor of quantized resistance $h/e^2$. The two-terminal resistance of a device therefore ensues from the parallel resistance of both edges, each of them being the sum of contributions of each helical edge sections. As a result, 
 \begin{equation}
 R_{\text{\text{2t}}}=\frac{h}{e^2}\left( \frac{1}{\text{N}_{\text{L}}} + \frac{1}{\text{N}_{\text{R}}} \right)^{-1}, 
 \label{eq1}
 \end{equation} 
 where N$_{\text{L,R}}$ is the number of helical conductor sections for the left (L) and right (R) edge between the source and drain contacts~\cite{Young14}. By swapping the source and drain contacts for various configurations of N$_{\text{L,R}}$, one expects to observe resistance plateaus given by Eq. (\ref{eq1}). Figure~\ref{Fig3}A displays a set of  four different configurations of two-terminal resistances measured at $B=2.5$~T as a function of back-gate voltage. Swapping the source and drain contacts and the number of helical edge sections (see contact configurations in Fig.~\ref{Fig3}B) yields a maximum around charge neutrality which reaches the expected values indicated by the dotted horizontal lines, thereby demonstrating helical edge transport. Notice that the plateau at $\frac{h}{e^2}$ in Fig.~\ref{Fig2}A is also fully consistent with Eq. (\ref{eq1}) for $\text{N}_{\text{L}}=\text{N}_{\text{R}}=2$ . 
 
 Four-terminal non-local configuration provides another stark indicator for helical edge transport~\cite{Roth09}. Figure~\ref{Fig3}C shows simultaneous measurements of the two-terminal resistance between the two blue contacts (see sample schematics in the inset), and the non-local resistance $R_{\text{NL}}$ measured on the red contacts while keeping the same source and drain current-injection contacts. While $R_{\text{2t}}$ nearly reaches the expected value  indicated by the dashed lines, namely $\frac{5}{6}\frac{h}{e^2}$ ($\text{N}_{\text{L}}= 5$ and $\text{N}_{\text{R}}=1$), a non-local voltage develops in the $V_{\text{bg}}$ range that coincides with the helical edge transport regime in $R_{\text{2t}}$. The large value of this non-local signal, much larger than what could be expected in the diffusive regime, demonstrates that current is flowing on the edges of the sample. For helical edge transport, the expected non-local resistance is given by $R_{\text{NL}}=R_{\text{2t}}\frac{\text{N}_{\text{V}}} {\text{N}_{\text{I}}}$, where N$_\text{I}$ and N$_\text{V}$ are the number of helical conductor sections between the source and drain contacts along the edge of the non-local voltage probes and between the non-local voltage probes, respectively. The measured  $R_{\text{NL}}$ shown in Fig.~\ref{Fig3}C is in excellent agreement with the expected value $\frac{1}{6}\frac{h}{e^2}$ (N$_\text{I} = 5$ and N$_\text{V}=1$) indicated by the dashed line. 

   \begin{figure*}[t!]
   	\includegraphics[width=1\linewidth]{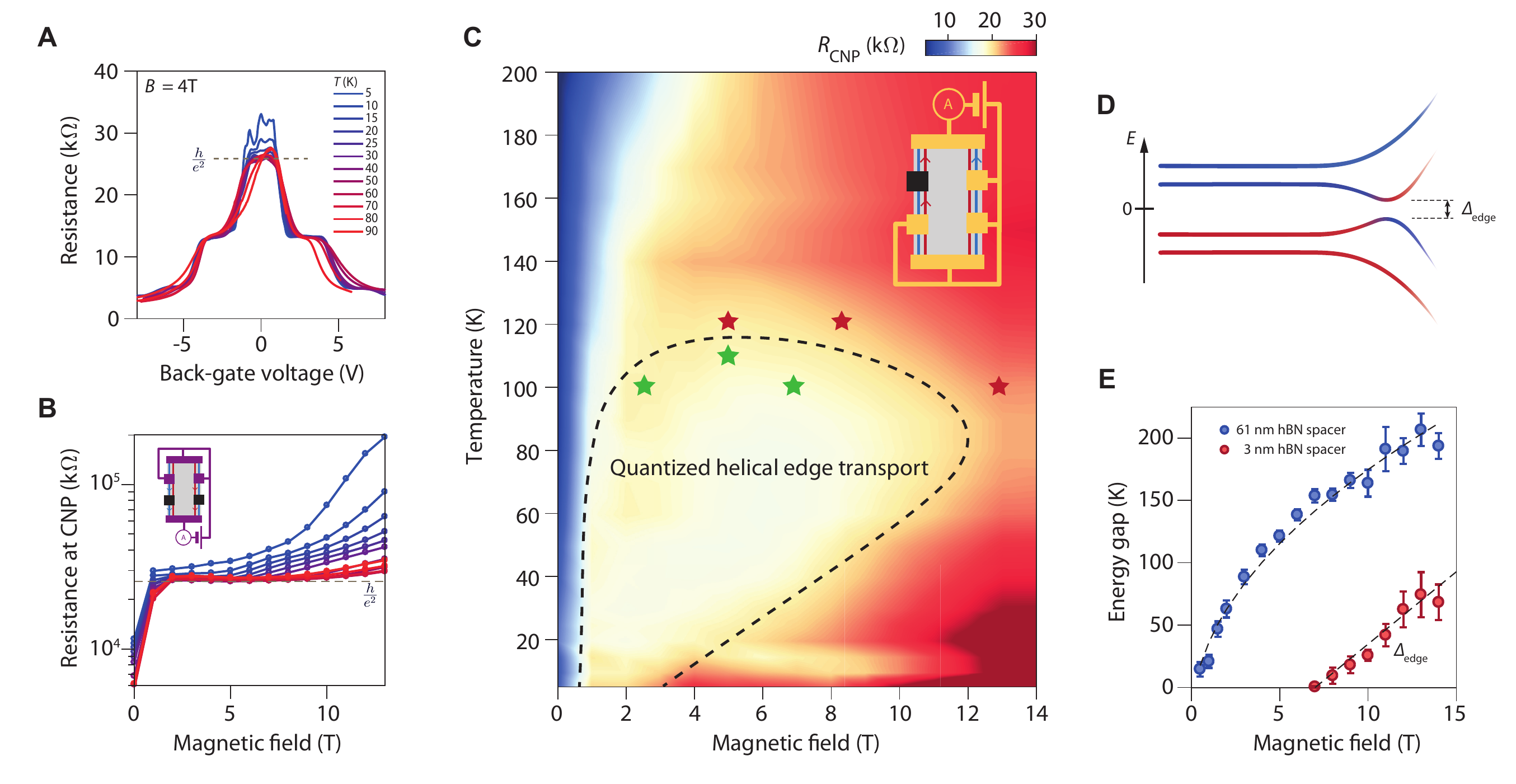} 
   	\caption{\textbf{Phase diagram of the helical edge transport.} (\textbf{A}) Two-terminal resistance of sample BNGrSTO-07 versus back-gate voltage measured at different temperatures and a magnetic field of $4$~T. The back-gate voltage is renormalized to compensate the temperature-dependence of the substrate dielectric constant (see Fig.~S12). (\textbf{B}) Two-terminal resistance at the charge neutrality point (CNP) for the same data as in (A). The inset shows the contact configuration used in (A) and (B). (\textbf{C}) Two-terminal resistance at the CNP versus magnetic field and temperature for a different contact configuration shown in the inset. The resistance shows a plateau at the value expected for helical edge transport ($\frac{2}{3}\frac{h}{e^2}$, color-coded in light yellow) over a large range of temperatures and magnetic fields, that is, up to $T=110$~K at $B = 5$~T. The stars indicate the parameters at which helical edge transport has been checked by measuring different contact configurations (green stars indicate quantized helical edge transport, red stars for deviation to quantization at the CNP). The dotted curve is a guide for the eye showing the approximate limits of the quantized helical edge transport. (\textbf{D}) Schematics of the edge dispersion of the zeroth Landau level broken-symmetry states showing the opening of a gap at the edge. (\textbf{E}) Activation energy at the charge neutrality point versus magnetic field measured in samples BNGrSTOVH-02 (red dots) BNGrSTO-09 (blue dots), which have a hBN spacer of 3 and 61~nm, respectively. The dotted lines are a linear fit for BNGrSTOVH-02 and a fit of the dependence $\alpha \sqrt{B} - \Gamma$  for BNGrSTO-09. The prefactor  $\alpha = 64$~KT$^{-1/2}$ corresponds to a disorder-free gap $\Delta = 0.4 \mathcal{E}_{\text{C}}$, and the intercept $\Gamma = 27$~K describes the disorder-broadening of the Landau levels, which is consistent with the sample mobility \cite{SM}.}
   	\label{Fig4}
   \end{figure*}
	
This global set of data that are reproduced on several samples \cite{SM} therefore provides compelling evidence for helical edge transport, substantiating the F phase as the ground state at charge neutrality of substrate-screened graphene.
 
 To assess the robustness of the helical edge transport we conducted systematic investigations of its temperature, $T$, and magnetic-field dependences. Figure~\ref{Fig4}C displays a color-map of the two-terminal resistance of sample BNGrSTO-07 measured at charge neutrality as a function of magnetic field and temperature. The expected resistance value for the contact configuration shown in the inset schematics is $R_{\text{2t}}=\frac{2}{3}\frac{h}{e^2}$. This quantized resistance value is matched over a remarkably wide range of temperature and magnetic field, which is delimited by the dashed black line, confirming the metallic character of the helical edge transport. To ascertain the limit of quantized helical edge transport we measured different contact configurations at some bordering magnetic field and temperature values (see Fig. S7), indicated in Fig.~\ref{Fig4}C by the green and red stars for quantized and not quantized resistance, respectively.  The temperature and magnetic field dependences are further illustrated by Fig.~\ref{Fig4}A and B, which show the two-terminal resistance (measured in a different contact configuration, see inset Fig.~\ref{Fig4}B) versus back-gate voltage and the resistance at the charge neutrality point versus $B$, respectively, for various temperatures. 
 
 These data show that quantized helical edge transport withstands very high temperatures, up to $110$~K, with an onset at $B\sim 1$~T virtually constant in temperature. Such a broad temperature range is comparable to WTe$_2$ for which QSH effect was observed in 100~nm-short channels up to $100$~K ~\cite{wu18}. The new aspect of the F phase of graphene is that the helical edge channels formed by the broken-symmetry states retain their topological protection over large distances at elevated temperatures, namely $1.1\,\mu$m for the helical edge section between contacts of the sample measured in Fig.~\ref{Fig4}A-C. 
 Various mechanisms can account for the high temperature breakdown of the helical edge transport quantization, such as activation of bulk charge carriers or inelastic scattering processes that break the U(1) spin-symmetry of the QHTI~\cite{Kharitonov16}. As the former would reduce resistance by opening conducting bulk channels, the upward resistance deviation upon increasing $T$ rather points to inelastic processes that do not conserve spin-polarization. Consequently, this suggests that quantized helical edge transport may be retained at even higher temperatures for lengths below $1\,\mu$m.
 
 Interestingly, the high magnetic field limit in Fig.~\ref{Fig4}C is temperature dependent. The lower the temperature the earlier in magnetic field start deviations from quantization: At $T=4$~K, we observe an increase of resistance on increasing $B$ from about $3$~T (see Fig.~\ref{Fig2}A and C and Fig.~\ref{Fig4}A-C), whereas this limit moves to $11$~T at $T=80$~K. For $B\gtrsim 3$~T, the resistance exhibits an activated insulating increase with lowering temperature, with a corresponding activation energy that increases linearly with $B$ (see Fig.~\ref{Fig4}E, red points. Data are taken on a different sample exhibiting an onset to insulation at $B\simeq 7$~T). Such a behaviour indicates a gap opening in the edge excitation spectrum as illustrated in the Fig.~\ref{Fig4}D schematics, breaking down the helical edge transport at low temperature. This linear $B$-dependence of the activation energy further correlates with the high magnetic field limit of the helical edge transport in Fig.~\ref{Fig4}C, thereby explaining why the limit for quantized helical edge transport increases to higher magnetic field with $T$.
 
 The origin of the gap in the edge excitation spectrum is most likely rooted in the enhancement of correlations with magnetic field. 
 An interaction-induced topological quantum phase transition from the QHTI to one of the possible insulating, topologically trivial quantum Hall ground states with spin or charge density wave order is a possible scenario~\cite{Kharitonov16}. Such a transition is expected to occur without closing the bulk gap~\cite{Young14,Kharitonov16}, which we confirmed via bulk transport measurements performed in a Corbino geometry (see Fig.~S8). Yet the continuous transition involves complex spin and isospin textures at the edges due to the $B$-enhanced isospin anisotropy~\cite{Kharitonov12b}, yielding the edge gap detected in Hall bar geometry. Another scenario relies on the helical Luttinger liquid~\cite{Wu06} behaviour of the edge channels, for which a delicate interplay between $B$-enhanced correlations, disorder and coupling to bulk charge-neutral excitations may also yield activated insulating transport~\cite{Tikhonov16}. 
 
 To firmly demonstrate the key role of the SrTiO$_3$ dielectric substrate in the establishment of the F phase, we conducted identical measurements on a sample made with a $60$~nm thick hBN spacer, much thicker than $l_B$ at the relevant magnetic fields of this study, so that screening by the substrate is irrelevant in the quantum Hall regime. Shown in Fig.~\ref{Fig2}C with the blue dots, the resistance at the charge neutrality point diverges strongly upon applying a small magnetic field, thus clearly indicating an insulating ground state without edge transport. Systematic study of the activated insulating behaviour leads to an activation gap that grows as $\sqrt{B}$ (see blue dots in Fig.~\ref{Fig4}E and Fig. S9), as expected for a charge excitation gap that scales as the Coulomb energy $\mathcal{E}_{\text{C}}=e^2/4\pi\epsilon_0\epsilon_{\text{BN}}l_B$ where $\epsilon_0$ and $\epsilon_{\text{BN}}$ are the vacuum permittivity and the relative permittivity of hBN. This self-consistently demonstrates that the F phase emerges as a ground state due to a significant reduction of the electron-electron interactions by the high-$k$ dielectric environment. 
 
 Understanding the substrate-induced screening effect for our sample geometry requires electrostatic considerations that take into account the ultra-thin hBN spacer between the graphene and the substrate \cite{SM}. The resulting substrate-screened Coulomb energy scale $\tilde{\mathcal{E}}_{\text{C}}=\mathcal{E}_{\text{C}} \times S(B)$ is suppressed by a screening factor $S(B)= 1-\frac{\epsilon_{\text{STO}}-\epsilon_{\text{BN}}}{\epsilon_{\text{STO}}+\epsilon_{\text{BN}}}\frac{l_B}{\sqrt{l_B^2+4d_{\text{BN}}^2}}$ where $\epsilon_{\text{STO}}$ is the relative permittivity of SrTiO$_3$. As a result, electrons in the graphene plane are subject to an unusual $B$-dependent screening that depends on the ratio $l_{B}/d_{\text{BN}}$ and is most efficient at low magnetic field (Fig. S11). Importantly, despite the huge dielectric constant of SrTiO$_3$ of the order of $\epsilon_{\text{STO}} \approx 10^4$ (Fig. S3), $\tilde{\mathcal{E}}_{\text{C}}$ is scaled down by a factor 10 for $l_B / d_{\text{BN}}=4$ due to the hBN spacer, which is still a significant reduction of the long-range Coulomb interaction. 
 
 How such a screening affects the short-range, lattice-scale contributions of the Coulomb and electron-phonon interactions that eventually determine the energetically favorable ground state is a challenging question since the hBN spacer precludes screening at the lattice scale. While theoretical estimates of these anisotropy terms initially point to the spin-polarized F ground state, renormalization effects due to the long-range part of the Coulomb interaction~\cite{Aleiner07,Basko08} result in anisotropy terms that can change signs and amplitudes in an unpredictable fashion~\cite{Kharitonov12}. The insulating behaviour commonly observed in usual samples still points to a canted anti-ferromagnetic order at charge neutrality, bearing out the strong renormalization of the anisotropy terms. We conjecture that the reduction of the Coulomb energy scale by the substrate screening is the key that suppresses the renormalization effects, restoring the F phase as the ground state at charge neutrality. Therefore, enhancing the Coulomb energy scale $\tilde{\mathcal{E}}_{\text{C}}$ by decreasing the ratio $l_{B}/d_{\text{BN}}$ induces a topological quantum phase transition from the QHTI ferromagnetic phase to an insulating, trivial quantum Hall ground state.
 
 Finally, our work demonstrates that the F phase in screened graphene, which emerges at low magnetic field, provides a prototypical, interaction-induced topological phase, exhibiting remarkably robust helical edge transport in a wide parameter range. The role of correlations in the edge excitations, which are tunable via the magnetic field and an unusual $B$-dependent screening, should be of fundamental interest for studies of zero-energy modes in superconductivity-proximitized architectures constructed on the basis of helical edge states~\cite{Fu09,Zhang14,SanJose15}. We further expect that substrate-screening engineering, tunable via the hBN spacer thickness, could have implications for many other correlated 2D systems for which the dielectric environment drastically impact their ground states and (opto)electronic properties.

 \section*{Acknowledgments}
 
 We thank H. Courtois, M. Goerbig, M. Kharitonov, A. MacDonald, and A. Grushin for valuable discussions. Samples were prepared at the Nanofab facility of N\'{e}el Institute. This work was supported by the H2020 ERC grant \textit{QUEST} No. 637815. K.W. and T.T. acknowledge support from the Elemental Strategy Initiative conducted by the MEXT, Japan, A3 Foresight by JSPS and the CREST (JPMJCR15F3), JST. Z.H. acknowledges the support from the National Key R\&D Program of China (2017YFA0206302), the National Natural Science Foundation of China (NSFC) with Grant 11504385, and the Program of State Key Laboratory of Quantum Optics and Quantum Optics Devices (No. KF201816). 
 
\bibliography{Graphene-QSH}


\clearpage
\setcounter{figure}{0}
\setcounter{section}{0}
\renewcommand{\thefigure}{S\arabic{figure}}


\begin{center}
\textbf{\large Supplementary Materials} \vspace{5mm}
\end{center}

	\section{I. Sample fabrication}
	
	hBN/graphene/hBN heterostructures were made from exfoliated flakes using the van der Waals pick-up technique \textit{(30)}. Contacts were patterned by  electron-beam lithography and metallized by e-gun evaporation of a Cr/Au bilayer after etching the stack directly through the resist pattern used to define the contacts. For sample BNGrSTOVH-02, a Hall-bar was first patterned by etching the stack using a CHF$_3$/O$_2$ plasma and a mask of hydrogen silsesquioxane (HSQ) resist. Contacts were then designed and deposited on the etched Hall-bar with apparent graphene edges. The design and contacting process of the sample in a Corbino geometry (BNGrSTO-Corbino1) is detailed in supplementary text section 5. For all samples we used 500$\,\mu$m thick SrTiO$_3$ (100) substrates that are cleaned with hydrofluoric acid buffer solution before deposition of the hBN/graphene/hBN heterostructures.

	\section{II. Measurements}
	
	Two and four-terminals measurements were performed with a standard low-frequency lock-in amplifier technique in voltage-bias configuration by applying an ac bias voltage of $0.4$~mV. To compensate for the dielectric hysteresis of the SrTiO$_3$ substrate, back-gate dependence of the resistance was systematically carried out with the same back-gate voltage sweep. This enabled to obtain reproducible position of the charge neutrality point in back-gate voltage. For the sake of clarity, the back-gate voltage axes of all figures are shifted to have the charge neutrality point at zero voltage.
	
\section{III. Samples summary}
\label{secSamples}

\begin{figure}[t!]
	\centering
	\includegraphics[width=0.8\linewidth]{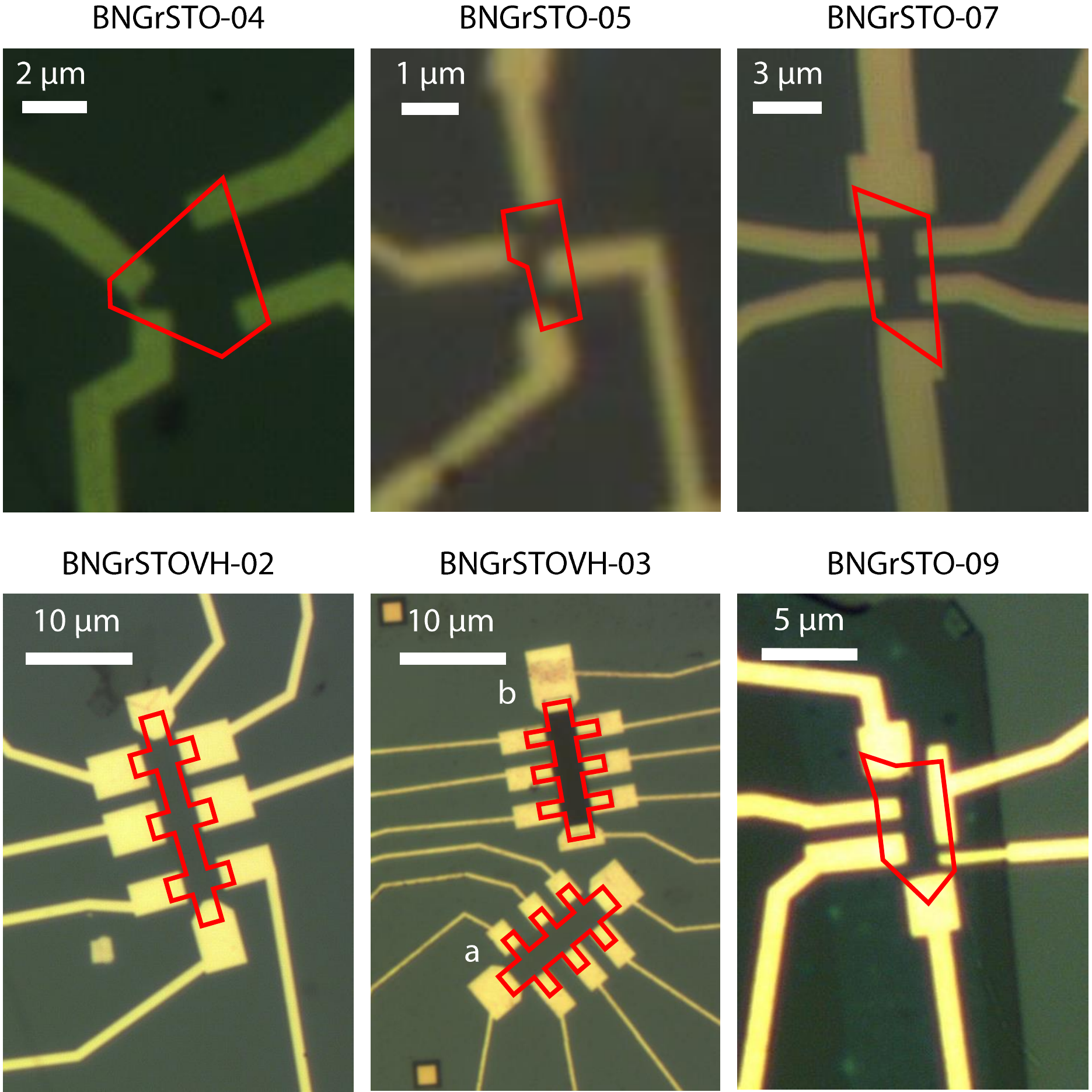}
	\caption{\textbf{Graphene devices.} Optical pictures of the samples listed in Table S1. The red lines underline the edges of the hBN-encapsulated graphene flakes. An optical picture of the sample in a Corbino geometry is shown in Fig. \ref{figS5}.}
	\label{figS0}
\end{figure}

\begin{figure}[h!]
	\centering
	\includegraphics[width=1\linewidth]{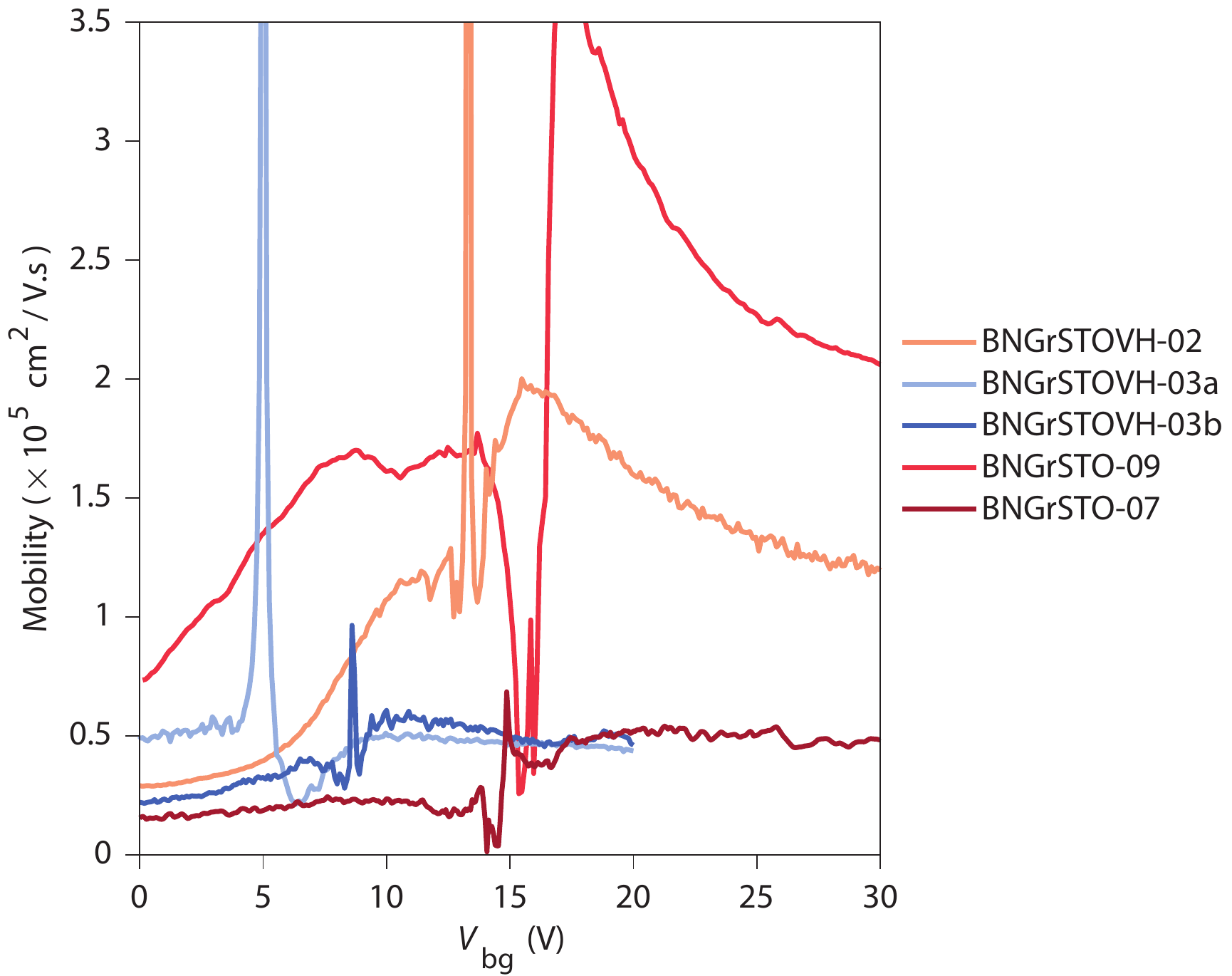}
	\caption{\textbf{Sample mobility.} Charge carrier mobility in five different Hall bars measured at $T=4\;$K.}
	\label{figS1a}
\end{figure}

In this section we present a summary of the different samples studied. Optical images of the samples are shown in Fig. \ref{figS0} and their geometry and transport characteristics are listed in Table S1. Thicknesses of the bottom hBN flakes for the samples showing an helical phase at low magnetic field vary between 1.6~nm and 5~nm. Graphene flake sizes vary from 1.6~$\mu$m up to 15~$\mu$m. The highest Hall mobility in our samples with thin hBN spacer reaches up to $130\,000$~cm$^2$/V.s for sample BNGrSTOVH-02, as shown in Fig.~\ref{figS1a}. 

Moreover, a sample with a thick bottom hBN layer was measured to confirm the importance of the hBN spacer thickness. Transport measurements of this sample are shown in supplementary text section 6. A device with a Corbino geometry was also studied to investigate bulk transport and the presence of a bulk gap at charge neutrality. Details about its fabrication and transport measurements are presented in supplementary text section 5. 

\begin{table}[h!]
	\begin{center}
		\begin{tabular}{|c|c|c|c|c|}
			\hline
			Sample & \begin{tabular}{@{}c@{}}Con- \\ tacts\end{tabular} & \begin{tabular}{@{}c@{}}Graphene \\ size ($\mu$m)\end{tabular} & \begin{tabular}{@{}c@{}}hBN spa- \\ cer (nm)\end{tabular} & \begin{tabular}{@{}c@{}}Mobility \\ (cm$^2$/V.s)\end{tabular} \\ \hline
			BNGrSTO-04 & 3  & $4 \times 5$ & 5 & $ 75\,000$ \\ \hline
			BNGrSTO-05 & 4  & $1.6 \times 2.2$  & 3.3 & $ 20\,000$ \\ \hline
			BNGrSTO-07 & 6 & $6 \times 3$ & 3.2 & $ 40\,000$ \\ \hline
			BNGrSTOVH-02 & 8 & $15 \times 2$ & 5 & $ 130\,000$ \\ \hline
			BNGrSTOVH-03a & 8 & $15 \times 3$ & 4 & $ 50\,000$ \\ \hline
			BNGrSTOVH-03b & 8 & $15 \times 3$ & 4 & $ 50\,000$ \\ \hline
			BNGrSTO-Corbino1 & 2 & $25 \times 10 $ & 1.6 & / \\ \hline
			BNGrSTO-09 & 6 & $8 \times 3 $ & 61 & $ 200\,000$ \\ \hline
		\end{tabular}
	\end{center}
	\caption{\textbf{Samples characteristics.} For samples BNGrSTO-04 and BNGrSTO-05, the mobility is extracted from the sheet resistance on the electron side at a back-gate voltage 10~V away from the Dirac point, and using an estimate of the geometrical capacitance of the SrTiO$_3$ substrate. For the other samples, the mobility is the Hall mobility measured at high electron density, see Fig.~\ref{figS1a}.}
\end{table}

\section{IV. Dielectric constant at \textit{T}$\mathbf{=4}$~K}

The doping of the hBN-encapsulated graphene flake with the application of a back-gate voltage is usually modeled by an equivalent plate capacitor. The corresponding capacitance $C_{\text{bg}}$ is the sum of the two capacitances coming from the hBN spacer and the SrTiO$_3$ (STO) substrate, such that:
\begin{equation}
\frac{1}{C_{\text{bg}}}=\frac{1}{C_{\text{STO}}}+\frac{1}{C_{\text{BN}}}\Rightarrow \frac{d_{\text{STO}}+d_{\text{BN}}}{\epsilon_r}=\frac{d_{\text{STO}}}{\epsilon_{\text{STO}}}+\frac{d_{\text{BN}}}{\epsilon_{\text{BN}}}
\end{equation}
where $d_{\text{STO}}$ and $d_{\text{BN}}$ are respectively the thicknesses of the SrTiO$_3$ substrate and the hBN spacer and $\epsilon_{\text{STO}}$, $\epsilon_{\text{BN}}$, $\epsilon_r$ are respectively the relative dielectric constants of the SrTiO$_3$ substrate, the hBN spacer and the overall system.
Since $d_{\text{BN}}\approx 1-5\;\text{nm}\ll d_{\text{STO}}=500\;\mu\text{m}$, the resulting relative permittivity $\epsilon_r$ is:
\begin{equation}
\epsilon_r=\frac{\epsilon_{\text{STO}}}{1+\frac{d_{\text{BN}}}{d_{\text{STO}}}\frac{\epsilon_{\text{STO}}}{\epsilon_{\text{BN}}}}
\end{equation}
At low temperature, the relative dielectric constant of the hBN spacer is $\epsilon_{\text{BN}}=4$. Thus with $\epsilon_{\text{STO}}\sim 10^4$, we obtain $d_{\text{BN}}\:\epsilon_{\text{STO}}/d_{\text{STO}}\:\epsilon_{\text{BN}}\lesssim 0.1$. The gate capacitance is therefore mainly determined by the relative permittivity of SrTiO$_3$ and one can assume that $\epsilon_{\text{STO}}\approx \epsilon_r$.  

\begin{figure}[h!]
	\centering
	\includegraphics[width=1\linewidth]{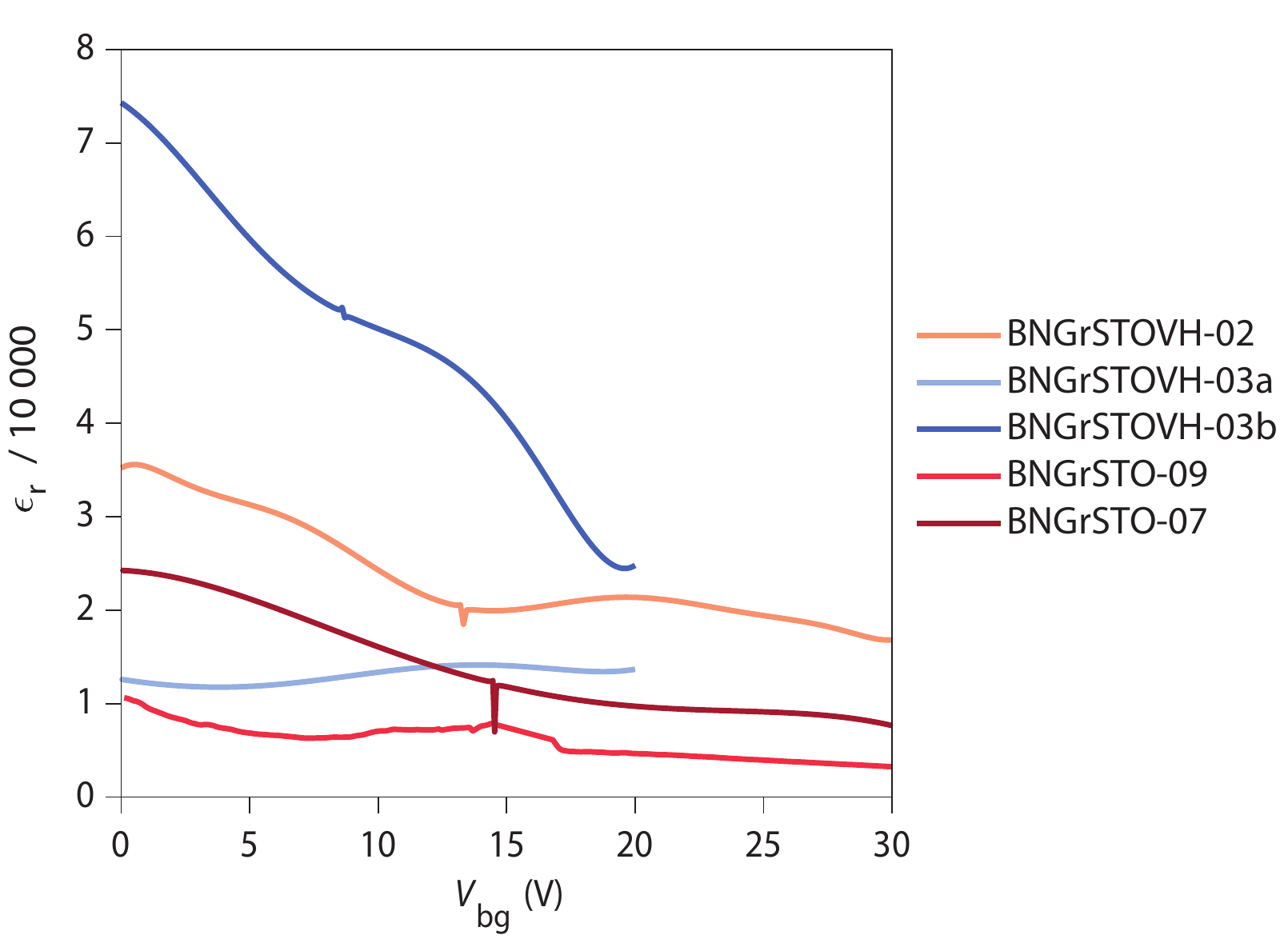}
	\caption{\textbf{Dielectric constant at \textit{T} $\mathbf{=4\;}$K.} Dielectric constant $\epsilon_r\approx \epsilon_{\text{STO}}$ versus back-gate voltage $V_{\text{bg}}$ for five different samples at $T=4$\;K. The dielectric constant is extracted from the graphene device charge carrier density obtained by Hall measurements.}
	\label{figS1}
\end{figure}

The  dielectric constant measured for different Hall bars at $T = 4\;$K is shown in Fig.~\ref{figS1}. $\epsilon_r$ is non-linear with the electric field and ranges from 5 000 to 35 000, coherent with the expected value for the dielectric constant of SrTiO$_3$ at low temperature. One substrate even shows a surprisingly high dielectric constant reaching 70 000 at $V_{\text{bg}} =0$~V.

\section{V. Helical edge transport in other devices}

In this section we present transport measurements performed on additional samples displaying characteristic signatures of the ferromagnetic (F) phase with helical edge states.
\begin{figure*}[tbp]
	\centering
	\includegraphics[width=0.8\linewidth]{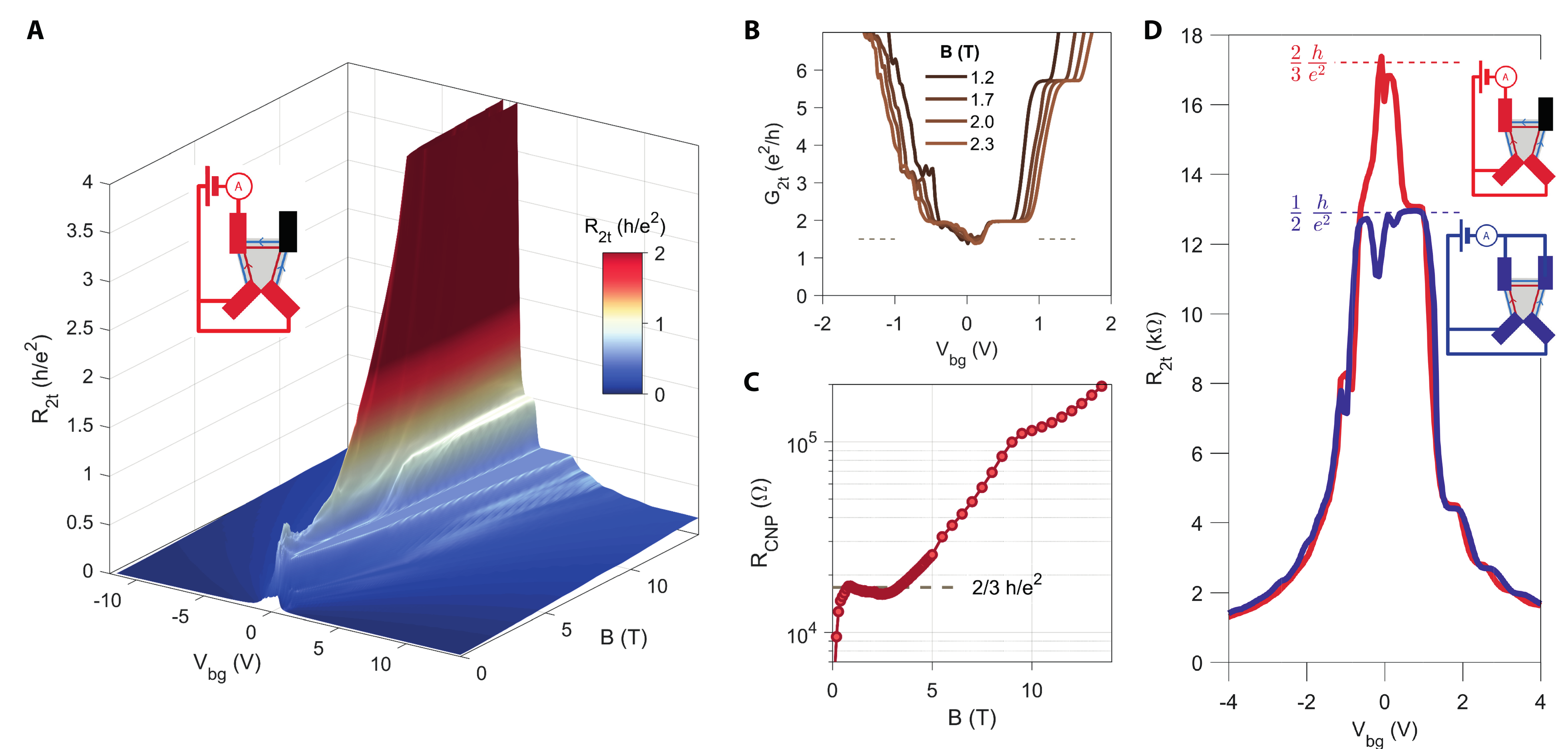}
	\caption{\textbf{Low magnetic field quantum spin Hall effect in sample BNGrSTO-04.} (\textbf{A}) Two-terminal resistance $R_{\text{2t}}$ in units of $h/e^2$ of sample BNGrSTO-04 versus magnetic field $B$ and back-gate voltage $V_{\text{bg}}$ measured at $4$~K. The inset schematics indicates the contact configuration. In addition to standard quantum Hall plateaus, the resistance exhibits an anomalous plateau around the charge neutrality point for magnetic field between $B=1$ and $3$~T which signals the regime of the quantum spin Hall effect in this sample. The value of the resistance at this plateau is $\frac{2}{3}\:h/e^2$. (\textbf{B}) Two-terminal conductance $G_{\text{2t}}=1/R_{\text{2t}}$ in units of $e^2/h$ versus back-gate voltage $V_{\text{bg}}$ extracted from (A) at different magnetic fields chosen in the quantum spin Hall state. The first conductance plateaus of the quantum Hall effect at $2$ and $6\,e^2/h$ are well defined for electron doping. The QSH plateau of conductance at $\frac{3}{2}\:e^2/h$ clearly emerges at charge neutrality around $V_{\text{bg}}=0$~V and is indicated by the gray dotted line. \textbf{(C)} Resistance $R_{\text{CNP}}$ at the charge neutrality point $V_{\text{bg}}=0$ versus $B$ extracted from (A) showing the persistence of the helical plateau up to $\sim 3$~T followed by a resistance increase towards an insulating state. (\textbf{D}) Two-terminal resistance $R_{\text{2t}}$ versus back-gate voltage $V_{\text{bg}}$ at a magnetic field of $B = 1.5\;$T, for different configurations of source and drain contacts. The expected resistance for helical edge transport for the corresponding contact configuration is indicated by dashed lines of the corresponding color. The corresponding contact configuration is shown in inset.}
	\label{figS6}
\end{figure*}

Figure~\ref{figS6} presents results of transport measurements for sample BNGrSTO-04, in the same fashion as in Fig. 2 of the main text. This sample is a Hall bar with 3 contacts (the lower two being shortcut), which leads to specific quantization values. As shown in Fig. \ref{figS6}A and B, the QSH plateau develops under a magnetic field between 1~T and 3~T. The resistance value reaches the expected value for helical edge transport in this contact configuration, that is, $\frac{2}{3}\:h/e^2$. In Fig. \ref{figS6}D we further show two different contact configurations with swapping contacts. In both cases, the resistance at charge neutrality reaches the expected value for helical transport indicated by the dotted red and blue lines.

\begin{figure*}[tbp]
	\centering
	\includegraphics[width=0.55\linewidth]{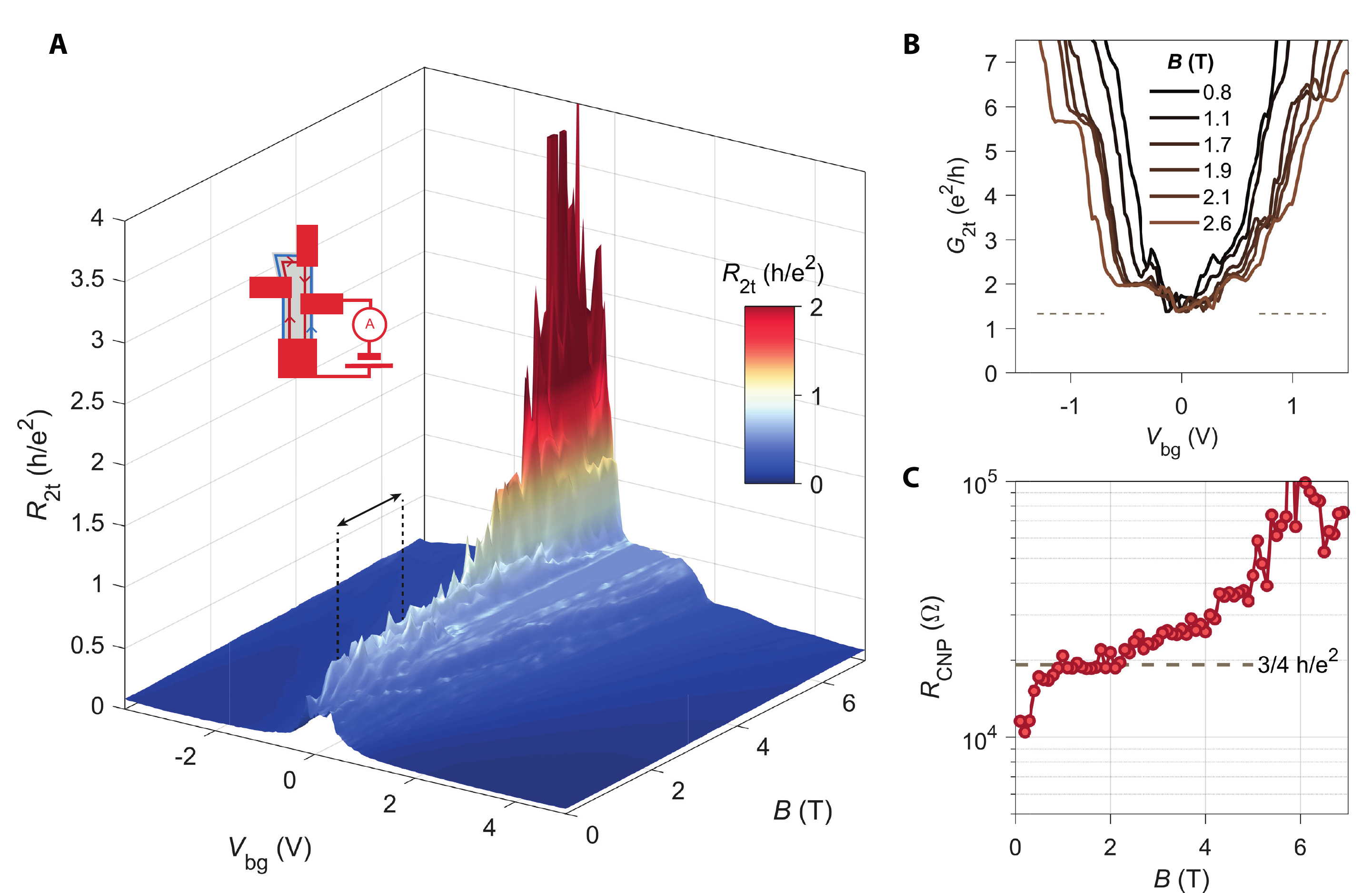}
	\caption{\textbf{Low magnetic field quantum spin Hall effect in sample BNGrSTO-05.} (\textbf{A}) Two-terminal resistance $R_{\text{2t}}$ in units of $h/e^2$ of sample BNGrSTO-05 versus magnetic field $B$ and back-gate voltage $V_{\text{bg}}$ measured at $4$~K. The inset schematics indicates the contact configuration. In addition to standard quantum Hall plateaus, the resistance exhibits an anomalous plateau around the charge neutrality point for magnetic field between $B=1$ and $2.5$~T, delimited by the black dotted lines and the arrow, which signals the regime of the quantum spin Hall effect in this sample. The value of the resistance at this plateau is $\frac{3}{4}\:h/e^2$. (\textbf{B}) Two-terminal conductance $G_{\text{2t}}=1/R_{\text{2t}}$ in units of $e^2/h$ versus $V_{\text{bg}}$ extracted from (A) at different magnetic fields chosen in the quantum spin Hall state. The first conductance plateaus of the quantum Hall effect at $2$ and $6\,e^2/h$ are well defined for hole doping. The QSH plateau of conductance at $\frac{4}{3}\,e^2/h$ clearly emerges at charge neutrality around $V_{\text{bg}}=0$~V and is indicated by the gray dotted line. (\textbf{C}) Resistance $R_{\text{CNP}}$ at the charge neutrality point $V_{\text{bg}}=0$ versus $B$ extracted from (A) showing the persistence of the helical plateau up to $\sim 2.5$~T followed by a resistance increase towards an insulating state.}
	\label{figS7}
\end{figure*}

Figure~\ref{figS7} presents the data obtained on sample BNGrSTO-05 equipped with four contacts. We observe a resistance plateau at charge neutrality between 1~T and 2.6~T (see Fig. \ref{figS7}B and C), consistent with helical edge transport. 

\begin{figure*}[tbp]
	\centering
	\includegraphics[width=0.9\linewidth]{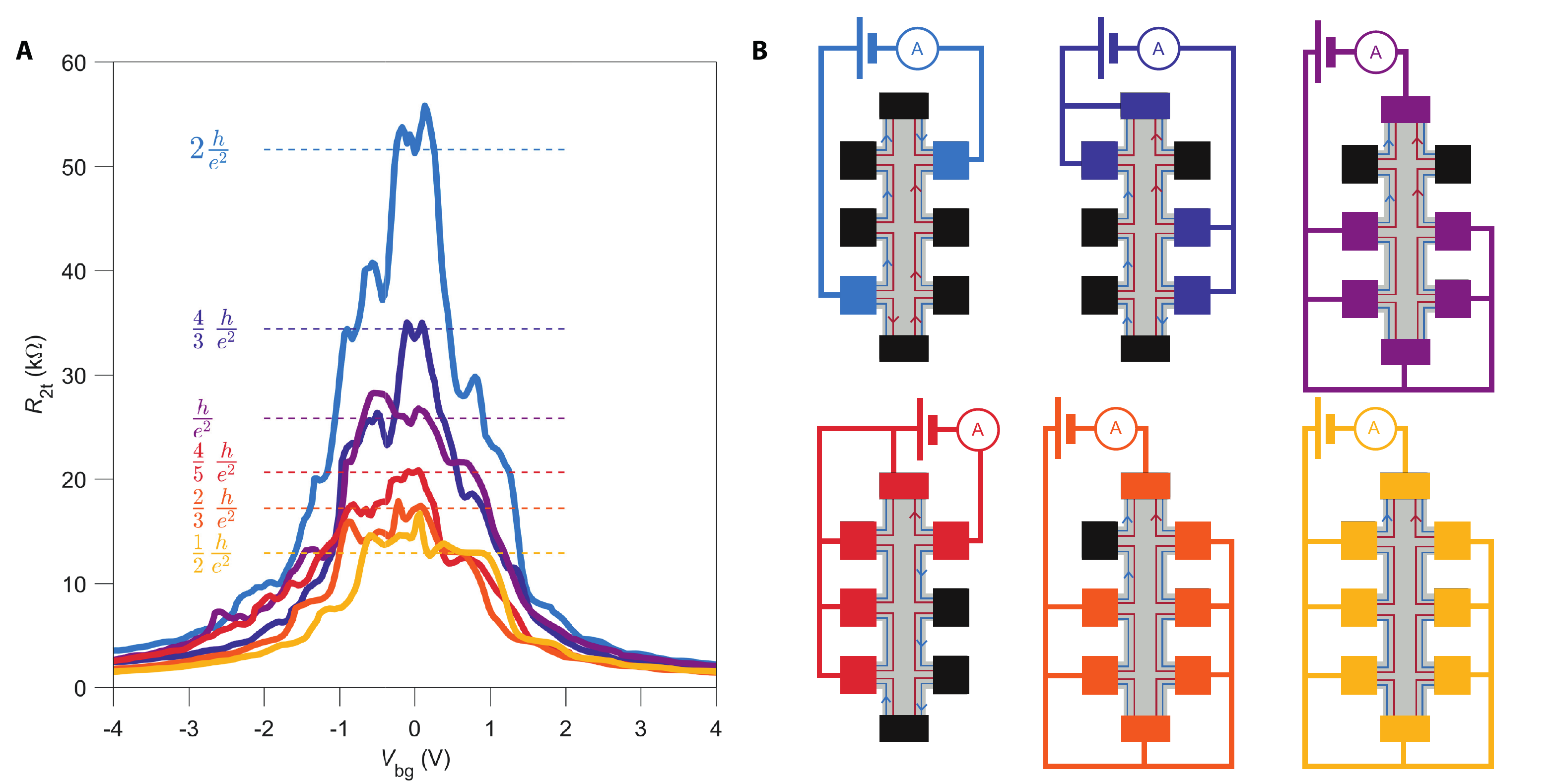}
	\caption{\textbf{Non-local helical edge transport in sample BNGrSTOVH-02.} (\textbf{A}) Two-terminal resistance $R_{\text{2t}}$ at $B = 1.5$~T and $T = 4$~K versus back-gate voltage $V_{\text{bg}}$, in sample BNGrSTOVH-02 for different contact configurations, presented in (B). Each contact configuration yields a resistance around charge neutrality reaching the expected values for helical edge transport, indicated by the horizontal dashed lines. (\textbf{B}) Schematics of the measurement configurations. Black contacts are floating. The red and blue arrows on the helical edge channels indicates the direction of the current between contacts.}
	\label{figS4}
\end{figure*}

In Figure~\ref{figS4} we present the evolution of the two-terminal resistance with respect to contact configuration for sample BNGrSTOVH-02, which is patterned in Hall bar geometry with eight contacts (see Fig.~\ref{figS0}). As in Fig.~2 of the main text, the resistance in six different configurations reaches around charge neutrality the quantized value expected for helical edge transport. 

Contrary to the sample BNGrSTO-07 presented in the main text, this sample was etched, so that the graphene edges are not native from the exfoliation but rough from plasma etching. This shows the independence of the F phase with respect to the edge roughness induced by plasma etching. The quantized value of $2\:h/e^2$ is reached over the complete length of the Hall bar, achieving a combined length of 15$\;\mu$m on each arm. Although the presence of floating contacts does not allow to draw conclusions on the coherence length, we can notice that disorder and edge roughness accumulated over $15\;\mu$m does not affect the quantization of the resistance in the F phase.

\section{VI. Resistance quantization for different points of the (\textbf{\textit{B}},\textbf{\textit{T}}) phase diagram}

\begin{figure*}[tbp]
	\centering
	\includegraphics[width=0.97\linewidth]{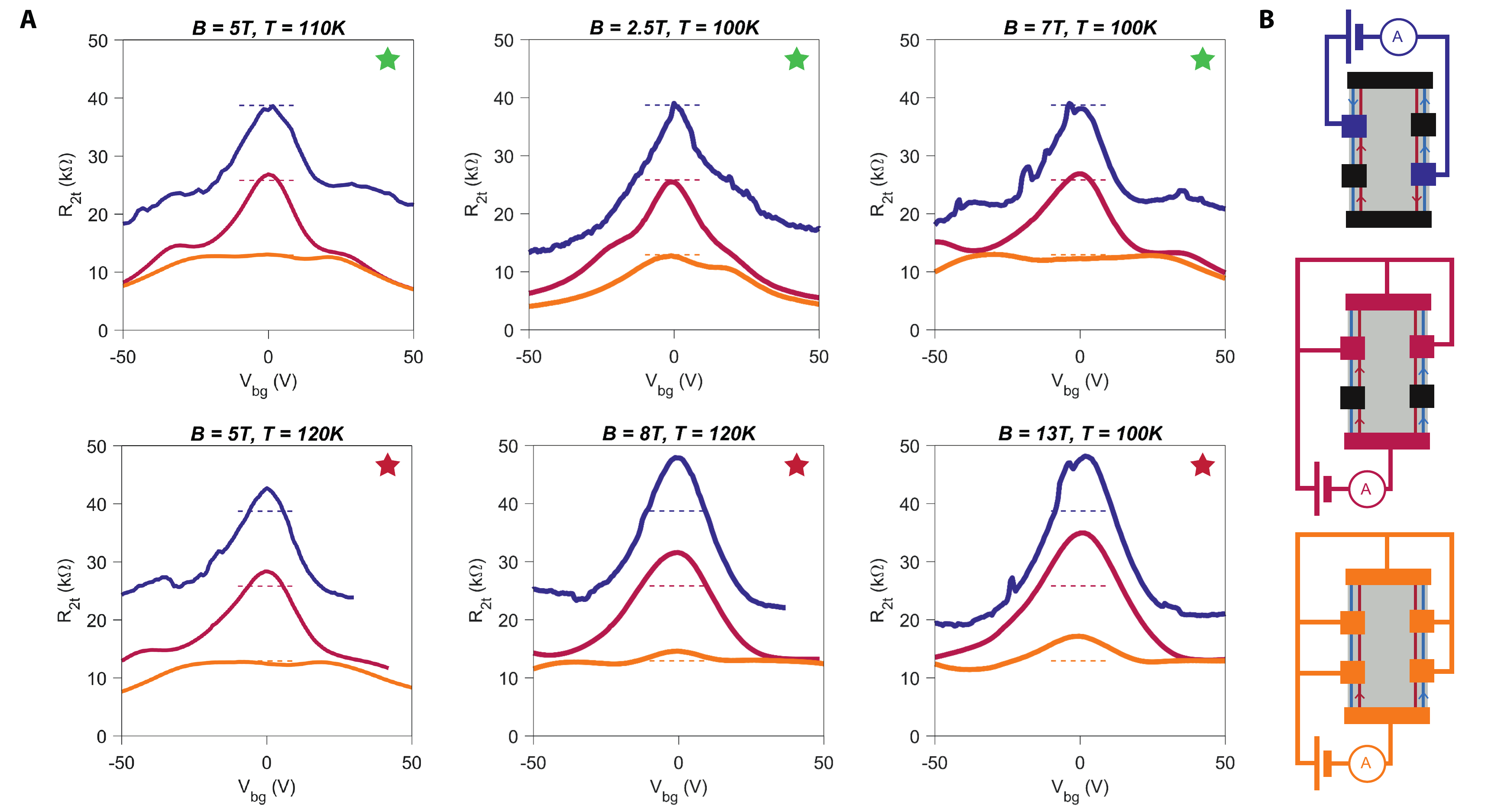}
	\caption{\textbf{Resistance quantization at different points in the (\textbf{\textit{B}},\textbf{\textit{T}}) phase diagram.} (\textbf{A}) Two-terminal longitudinal resistance $R_{\text{2t}}$ versus gate voltage $V_{\text{bg}}$ of sample BNGrSTO-07. Top (resp. bottom) row: points in the phase diagram corresponding to where the quantized helical transport (resp. loss of the quantization) is observed, labeled by green (resp. red) stars in Fig. 4C of the main text. (\textbf{B}) Corresponding contact configurations.}
	\label{figS10}
\end{figure*}

In this section we present the two-terminal resistance of sample BNGrSTO-07 measured at several temperatures and for different magnetic fields, for different contact configurations corresponding to the points marked by green and red stars in Fig.~4C of the main text. In Figure~\ref{figS10}A the top row plots show the expected quantization of the resistance at charge neutrality for the three contact configurations shown in Fig.~\ref{figS10}B. On the contrary, when passing the limit of the F phase as shown in Fig. 4C of the main text, all resistances increase beyond the expected helical value (see bottom row plots in Fig.~\ref{figS10}A). 

\section{VII. Gap opening measured in Corbino geometry}
\label{secCorbino}

Due to the presence of helical edge channels in the F phase, the existence of a bulk gap in the $\nu = 0$ state can not be inferred directly from transport measurements in a Hall bar geometry. Therefore to show the existence of a bulk band-gap at $\nu = 0$, we designed a graphene device in a Corbino geometry. In this geometry, graphene is contacted in the middle of the flake, preventing any edge states from connecting the different contacts, and giving access to bulk transport via measurement of the two-probe resistance between contacts.

\begin{figure*}[tbp]
	\centering
	\includegraphics[width=0.97\linewidth]{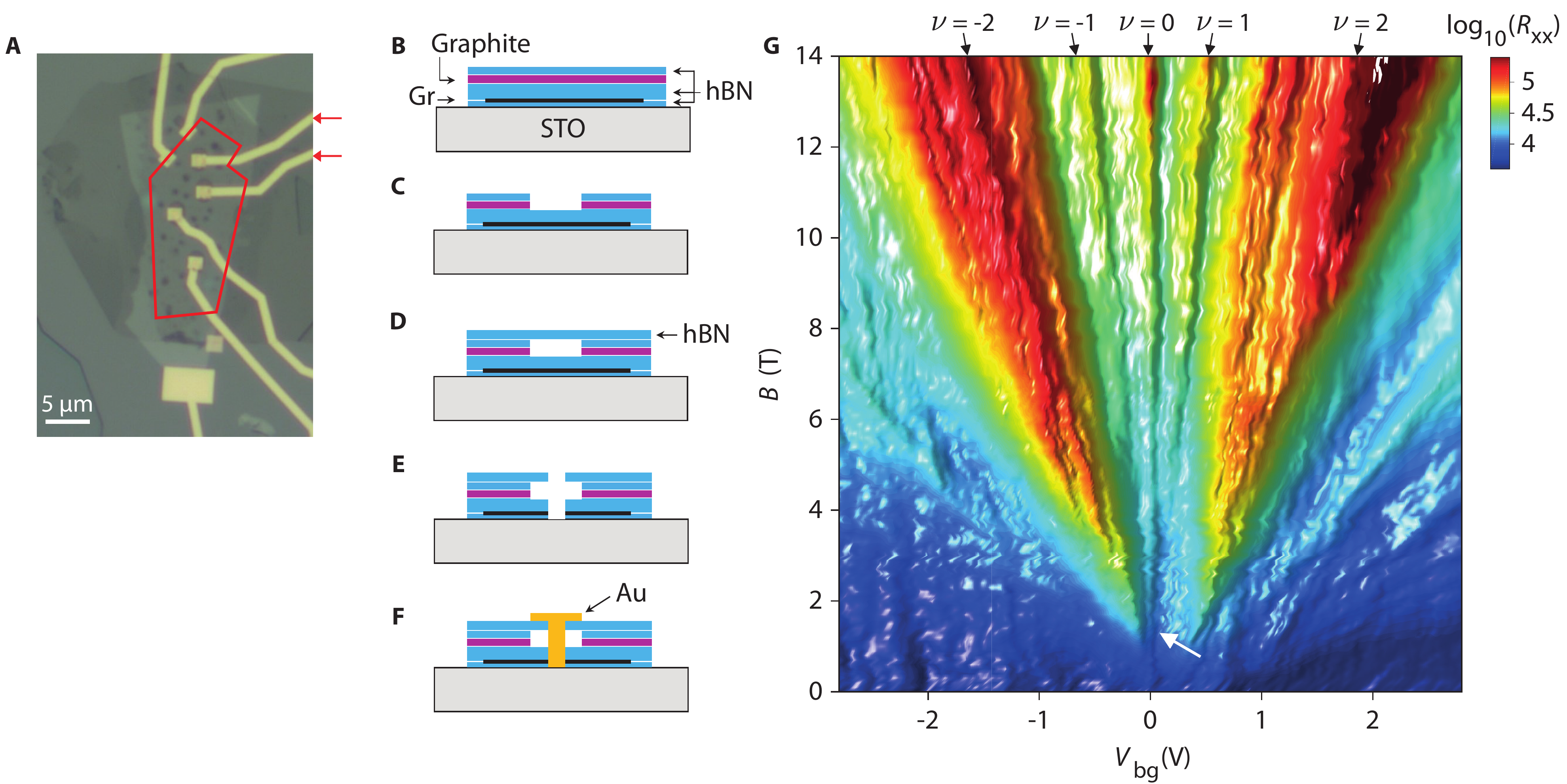}
	\caption{\textbf{Transport in a Corbino geometry.} (\textbf{A}) Optical image of the Corbino nanostructure. The graphene edges are underlined by the red line. The contacts used for transport are indicated by the red arrows. (\textbf{B-F}) Nanofabrication processing of a graphene heterostructure for the Corbino geometry. The top graphite layer of the original stack (\textbf{B}) is first etched in a large window (\textbf{C}). A thin hBN flake is then deposited on top of the stack (\textbf{D}). Finally, the whole stack is etched in the middle of the large window (\textbf{E}) to uncover the graphene edge, which is then contacted with Cr/Au contacts (\textbf{F}). (\textbf{G}) Fan diagram of the two-probe longitudinal resistance. The peak corresponding to the $\nu = 0$ gap is visible from $B = 1.5\;$T onward (onset marked by the white arrow).}
	\label{figS5}
\end{figure*}

An optical picture of our Corbino geometry device is shown in Fig.~\ref{figS5}A. The hBN/graphene/hBN stack was first placed on SrTiO$_3$ substrate, then covered with a graphite/hBN layer (see Fig.~\ref{figS5}B). The electrostatic screening of the graphite flake prevents the gold lines passing on the top of the heterostructure (see Fig. \ref{figS5}A) from doping locally the  graphene sample, which could induce quantum Hall edge states connecting the inner contacts to the graphene edges \cite{Polshyn2018,Zeng19}. The graphite layer was then etched to drill holes that enable to access and contact the graphene (see Fig.~\ref{figS5}C). A thin ($10\;$nm) hBN flake was deposited on top of the etched holes (Fig. \ref{figS5}D), to prevent contacting the graphite layer with the metallic electrodes. Smaller holes in this last hBN flake and the graphene flake were then made in the middle of the first etched holes via an additional etching step (Fig.~\ref{figS5}E). Finally, Cr/Au contacts were deposited in these holes to make edge contacts to the graphene (Fig.~\ref{figS5}F).

The device was measured in two-terminal configuration between the two inner contacts indicated with the red arrows in Fig. \ref{figS5}A. Fig. \ref{figS5}G shows a colormap of the two-terminal resistance, with respect to back-gate voltage and magnetic field. We observe resistance peaks, appearing at $ 1.5$~T around charge neutrality, which disperse with magnetic field, analogous to what can be observed in a Landau fan diagram. Those insulating peaks indicate the opening of gaps in the density of states between Landau levels and their broken-symmetry states. The two main resistance peaks arising in red color correspond to the cyclotron gaps between the $N=0$ and $N=\pm 1$ Landau levels, similarly to what was already observed in high mobility graphene devices on SiO$_2$ ~\cite{Polshyn2018, Zeng19}. Moreover, for $B > 1.5$~T, three resistance peaks develop near charge neutrality in the zeroth Landau level between the two main resistance peaks. 
A first central peak, not dispersing in magnetic field, appears above $1.5$~T at charge neutrality, which corresponds to the gap of the $\nu = 0$ state. Two additional satellites peaks appears above 5~T and correspond to the opening of the $\nu = \pm 1$ broken-symmetry states. 
Importantly, the gap at $\nu = 0$ state appears for a field value virtually identical to the one where helical edge transport at charge neutrality sets in in the different samples.  This bears out our conclusions of a F phase, gapped in the bulk and harboring helical edge channels, as the ground state of our screened graphene devices at charge neutrality.

\section{VIII. Activation gap at charge neutrality}
\label{secGap}

We present in this section the Arrhenius plots that lead to the activation gaps displayed in Figure 4E of the main text.

\begin{figure}[tbp]
	\centering
	\includegraphics[width=0.75\linewidth]{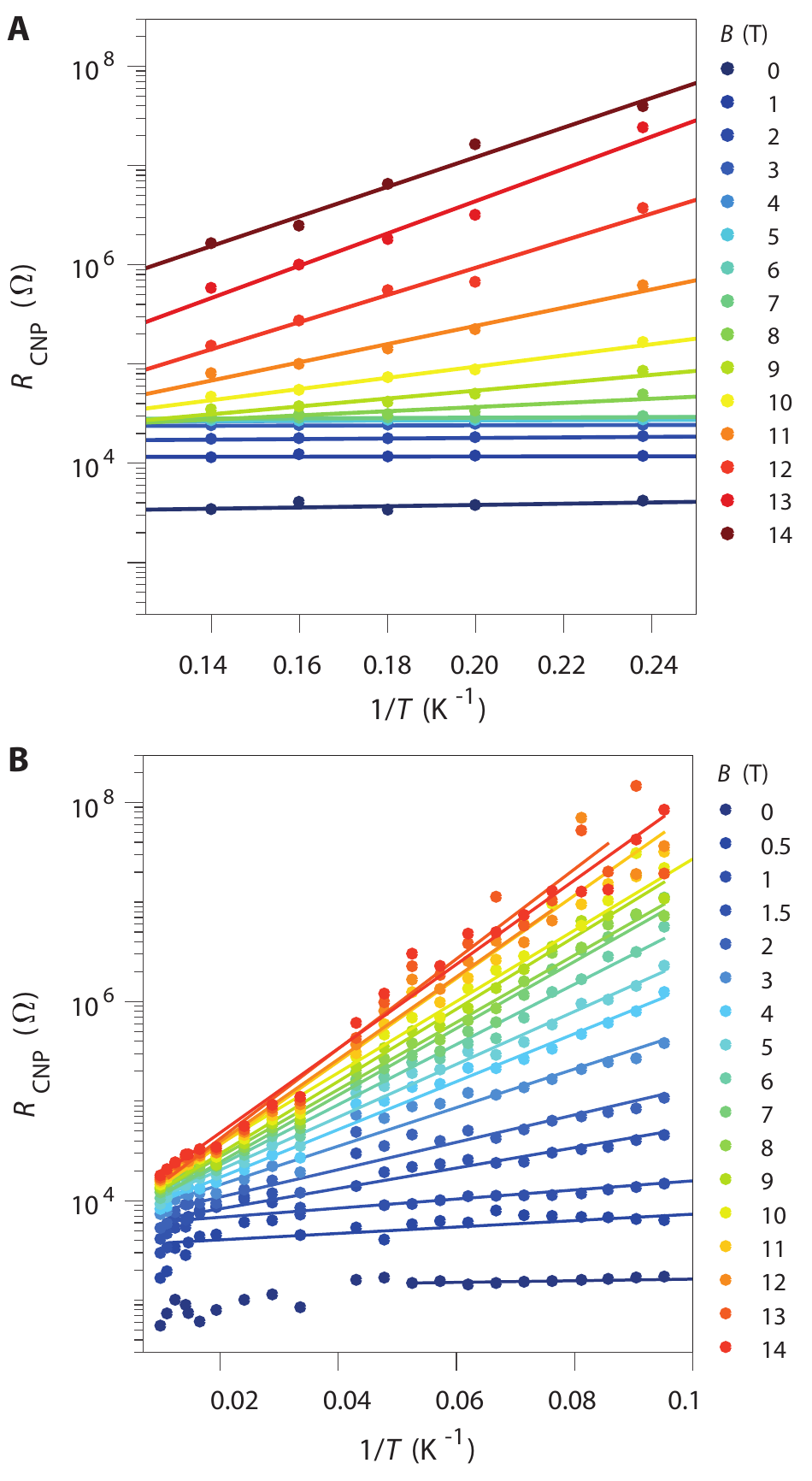}
	\caption{\textbf{Arrhenius plots.} Four-terminal longitudinal resistance at the charge neutrality point $R_{\text{CNP}}$ versus inverse temperature for different magnetic field values. Arrhenius plot measured for sample BNGrSTOVH-02 (\textbf{A}) and for sample BNGrSTO-09 (\textbf{B}).}
	\label{figS9}
\end{figure}

Figure~\ref{figS9}A shows the Arrhenius plot of the longitudinal four-terminal resistance measured in sample BNGrSTOVH-02, versus inverse temperature. This sample displays helical transport features (see Fig. \ref{figS4}).
For $B\gtrsim6$~T, we observe an activated temperature dependence of the resistance that we ascribe to the opening of a gap in the edge excitation spectrum. The $B$-evolution of this gap is shown in Fig.~4E of the main text. 

\begin{figure}[tbp]
	\centering
	\includegraphics[width=0.75\linewidth]{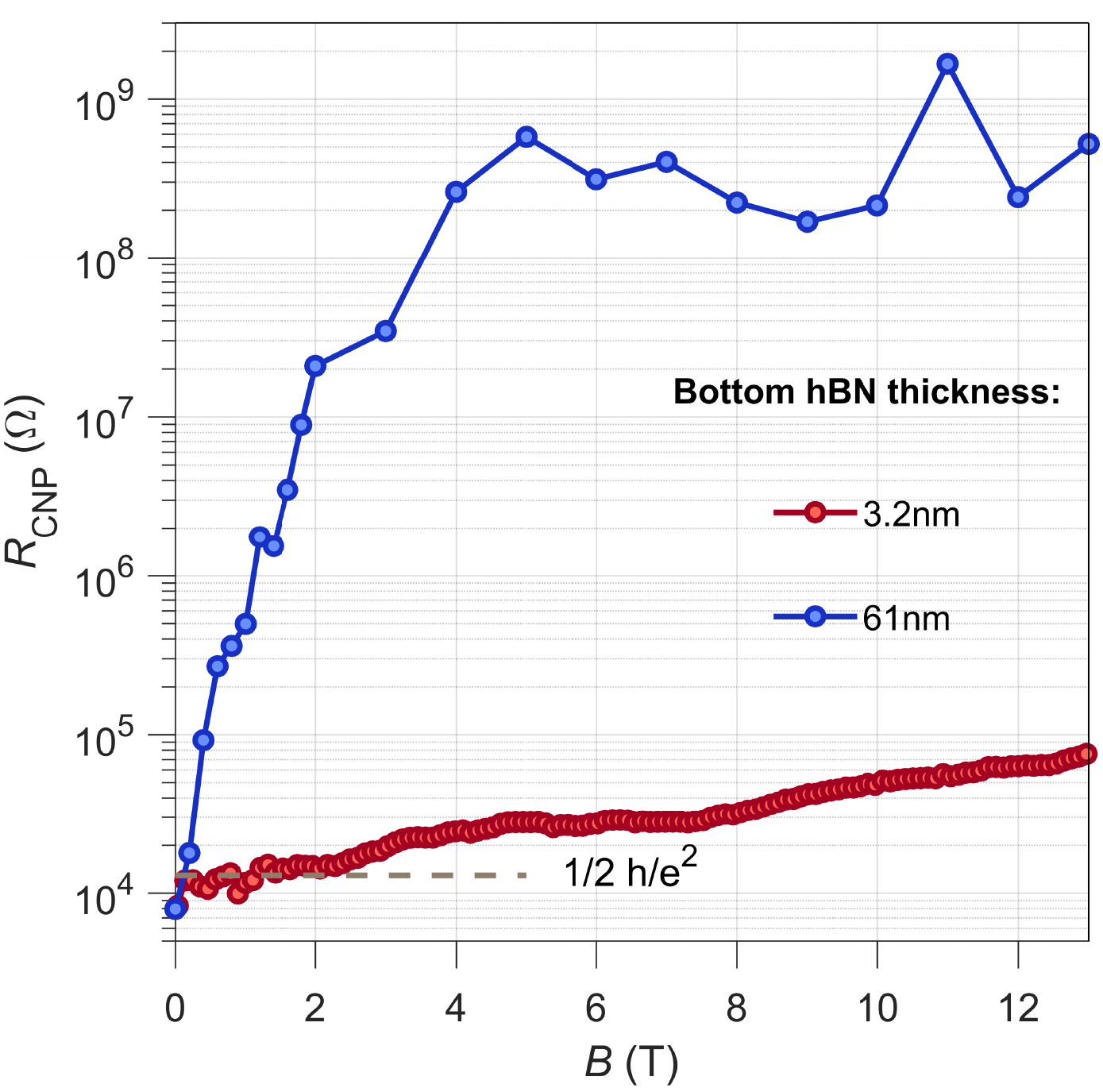}
	\caption{\textbf{Influence of the hBN spacer thickness.} Four-terminal longitudinal resistance at the charge neutrality point $R_{\text{CNP}}$ in sample BNGrSTO-07 (3.2~nm) and BNGrSTO-09 (61~nm) at $T=4$~K, versus magnetic field $B$. The expected resistance value at $\frac{1}{2}\:h/e^2$ for helical transport is marked by the grey dotted line.}
	\label{figS8}
\end{figure}

Figure~\ref{figS9}B presents the Arrhenius plot of the longitudinal four-terminal resistance measured in sample BNGrSTO-09. In this sample, the hBN spacer layer is 61~nm thick and no helical edge transport is observed.  We instead observe a strong insulating behaviour upon increasing magnetic field as seen on Fig.~\ref{figS8}, which shows the evolution of the longitudinal four-terminal resistance for both samples (notice that for sample BNGrSTO-07 the contact configuration is different from the one of Figure 2 of the main text). It emphasizes the fundamental difference of the insulating behaviours between samples with thin hBN spacer and samples with thick hBN spacer. Above 4~T, the saturation of the measured resistance for the sample with a 61~nm spacer corresponds to the noise level of our current amplifier.

\begin{figure*}[t!]
\centering
	\includegraphics[width=0.7\linewidth]{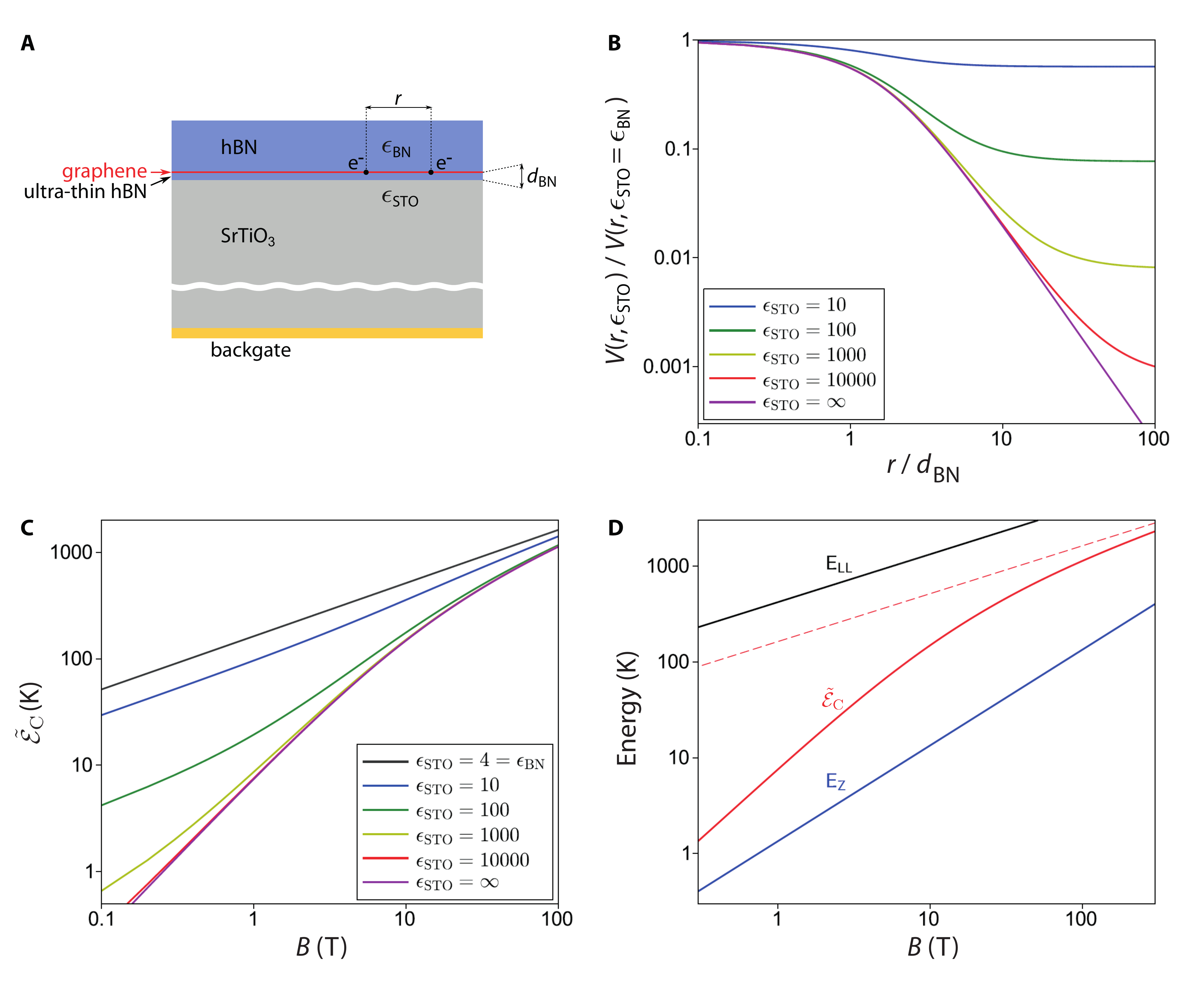}
	\caption{\textbf{Magnetic-field-dependent dielectric screening in graphene on SrTiO$\mathbf{_3}$ substrate.} (\textbf{A}) Cross-section of the hBN-encapsulated graphene heterostructure with an ultra-thin bottom hBN layer, deposited on a high-k dielectric SrTiO$_3$ substrate. The Coulomb interaction between graphene charge carriers is strongly reduced by the high dielectric permittivity of the substrate. (\textbf{B}) Screening factor in the Coulomb interaction plotted versus charge carrier separation for different substrate dielectric constants. The screening factor is significant for charge carrier separations $r$ larger than the bottom hBN layer thickness $d_{\rm BN}$. (\textbf{C}) Coulomb interaction $\tilde{\mathcal{E}}_{\text{C}}$ in the quantum Hall regime plotted versus magnetic field $B$ for different substrate dielectric constants and for a bottom hBN layer thickness $d_{\rm BN}=4$~nm. (\textbf{D}) Coulomb energy $\tilde{\mathcal{E}}_{\text{C}}$ and Zeeman energy $E_{\rm Z}$ compared to the cyclotron gap $E_{\rm LL}$ between the zeroth and first Landau levels. The Coulomb energy is large for substrates with low dielectric constant (red dashed line plotted for $\epsilon_{\rm STO}=\epsilon_{\rm BN}=4$), but is small at low magnetic field for substrates with high dielectric constant (red solid line plotted for $\epsilon_{\rm STO}=10\;000$ and $d_{\rm BN}=4$~nm).}
	\label{figS12}
\end{figure*}

\begin{figure*}[t!]
	\centering
	\includegraphics[width=0.8\linewidth]{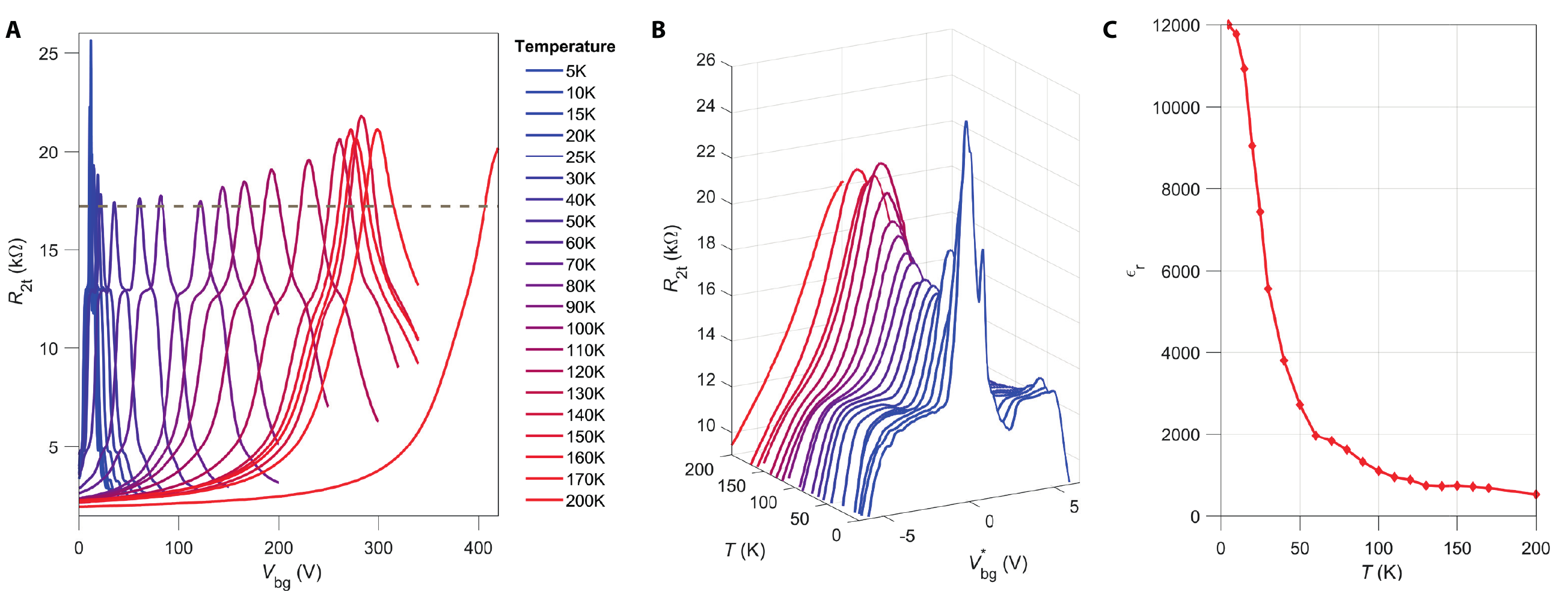}
	\caption{\textbf{Back-gate sweeps at different temperatures and extracted SrTiO$_3$ dielectric constant.} (\textbf{A}) Two-terminal longitudinal resistance $R_{\text{2t}}$ versus gate voltage $V_{\text{bg}}$ in sample BNGrSTO-07 at $B=5\;$T. From these data were extracted the resistances at charge neutrality presented in Fig. 4A of the main text. The dashed gray line indicates the expected resistance value for helical edge transport in this contact configuration. As the temperature increases, the dielectric constant of SrTiO$_3$ decreases and the back-gate effect becomes less effective, requiring higher voltages to induce the same carrier density differences. All sweeps start from zero gate voltage toward positive voltages. (\textbf{B}) Same data as in (A) but plotted against rescaled voltage $\textit{V}^*_{\text{bg}}$. (\textbf{C}) Dielectric constant $\epsilon_r\approx \epsilon_{\text{STO}}$ versus temperature $T$, as extracted from the two-terminal back-gate sweeps.}
	\label{figS11}
\end{figure*}

\section{IX. Magnetic-field-dependent dielectric screening in graphene \\ on \NoCaseChange{SrTiO$_3$} substrate}

In our van der Waals heterostructures, the dielectric screening in the graphene plane involves three different dielectric media : SrTiO$_3$, hBN, vacuum (see Fig. \ref{figS12}A). The presence of two dielectric interfaces gives rise to an infinite number of image charges, and the resulting expression contains an infinite number of terms~\cite{barcellona18}. Since the top hBN layer is much thicker than the bottom one and thicker than the magnetic length, one can greatly simplify the formula by considering that the top hBN layer is infinitely thick. In this case, the Coulomb interaction is given by the simple expression~\cite{Jackson75} :
\begin{equation}
V(r)=\frac{e^2}{4\pi\epsilon_0\,\epsilon_{\rm BN}\,r}\left(1-\frac{\epsilon_{\rm STO}-\epsilon_{\rm BN}}{\epsilon_{\rm STO}+\epsilon_{\rm BN}}\frac{r}{\sqrt{r^2+4\,{d_{\rm BN}}^2}}\right)
\label{eqCoulomb}
\end{equation}
where $r$ is the distance between the two electrons (or holes), $d_{\rm BN}$ is the thickness of the bottom hBN layer, $\epsilon_{\rm STO}\sim 10^4$ is the dielectric constant of the SrTiO$_3$ substrate, and $\epsilon_{\rm BN}=4$ the dielectric constant of the exfoliated top and bottom hBN flakes. The term in the parenthesis represents the additional screening brought by the SrTiO$_3$ substrate and depends on the distance between the charge carriers (see Fig. \ref{figS12}B). For separations much larger than the thickness of the bottom hBN layer, the Coulomb interaction is strongly reduced by the huge dielectric constant of the substrate, but for an intermediate distance $r/d_{\rm BN}=4$, the reduction is ``only'' by a factor 10.

In the quantum Hall regime, the distance $r$ between two electrons from the same Landau level is roughly given by the magnetic length $l_B=\sqrt{\hbar/eB}$ (26~nm at 1~T). Replacing $r$ by $l_B$ in the parenthesis of expression \eqref{eqCoulomb} gives a screening factor $S(B)$ which depends on the magnetic field. For a typical bottom hBN thickness of 4~nm in our heterostructures, the screening factor is about 0.008 at 1~T, 0.03 at 4~T, and 0.06 at 9~T. The characteristic Coulomb energy $\tilde{\mathcal{E}}_{\text{C}}$ can be written as : 
\begin{equation}
\tilde{\mathcal{E}}_{\text{C}} = \frac{e^2}{4\pi\epsilon_0\,\epsilon_{\rm BN}\,l_B} \times S(B)
\end{equation}
with the magnetic-field-dependent screening factor $S(B)$ given by:
\begin{equation}
S(B)=1-\frac{\epsilon_{\rm STO}-\epsilon_{\rm BN}}{\epsilon_{\rm STO}+\epsilon_{\rm BN}}\frac{l_B}{\sqrt{l_B^2+4\,{d_{\rm BN}}^2}}
\end{equation}
$S(B)$ reinforces the $1/l_B$ dependence of $\tilde{\mathcal{E}}_{\text{C}}$ (see Fig. \ref{figS12}C). As a result, the Coulomb energy varies much more with magnetic field with SrTiO$_3$ substrate than with usual SiO$_2$ substrates. The strong reduction of the Coulomb energy at low magnetic field makes it comparable to the Zeeman energy (see Fig. \ref{figS12}D).

\section{X. Temperature dependence of \NoCaseChange{SrTiO$_3$} dielectric constant}

The dielectric constant of SrTiO$_3$ exhibits a strong temperature dependence. It increases from a few hundreds at room temperature to about $10^4$ at low temperatures~\cite{Sakudo71}. As a result, the back-gate voltage one must apply to have a specific carrier density significantly increases with temperature due to the decrease of  $\epsilon_{\text{STO}}$ with $T$. Consequently, gate sweeps can only be performed up to a given temperature, fixed by the maximum back-gate voltage applicable. In our case, our maximal voltage limit of 420~V prevented us from carrying measurements above 200~K, where the charge neutrality point move to $418~$V at 14~T (see Fig.~\ref{figS11}A).

To illustrate in more details this effect, we present in Fig.~\ref{figS11} the two-terminal resistance measured as a function of the back-gate voltage for sample BNGrSTO-07 at $B=5\;$T, from which the data of Fig. 4C of the main text were extracted. As seen in Fig. \ref{figS11}A, the charge neutrality point moves to higher back-gate voltage values upon increasing $T$ due to the significant decrease of $\epsilon_{\text{STO}}$. Still, it is possible to rescale all back-gate voltage values into a equivalent back-gate voltage $\textit{V}^*_{\text{bg}}$:
\begin{equation}
\textit{V}^*_{\text{bg}} = \left(\textit{V}_{\text{bg}} - V_{\text{CNP}} (T)\right) \times \frac{\epsilon_r(T)}{\epsilon_r(T = 4\;\text{K})}
\end{equation}
where $\epsilon_r(T)$ is the temperature dependent dielectric constant of the back-gate electrode and $V_{\text{CNP}} (T)$ is the temperature dependent back-gate voltage at charge neutrality. When rescaled, the curves at different temperatures superimpose, as shown in Fig. \ref{figS11}B, up to a temperature above which the helical quantization at charge neutrality is not observed anymore, and the $\nu = \pm 2$ plateaus start to be smeared. 

We can furthermore use these data to estimate $\epsilon_r(T)$ by calculating the difference in gate voltage $\Delta V_{\text{bg}}$ between charge neutrality and the beginning of the $\nu=-2$ plateau at $B = 14$~T, which corresponds to a fixed carrier density, via the relation: 
$$
\frac{\Delta V_{\text{bg}}(T = 4\;\text{K})}{\Delta V_{\text{bg}}(T)} = \frac{\epsilon_r(T)}{\epsilon_r(T = 4\;\text{K})}
$$
For our sample, we determined by Hall measurements $\epsilon_r = 12~000$ at $V_{\text{bg}} = 15\;$V and $T=4.2\;$K.

The result of this analysis is shown in Fig. \ref{figS11}C. The dielectric constant rapidly decreases with temperature but is still close to 1000 at 200~K.

\end{document}